\documentclass[aps,prd,eqsecnum,11pt,showpacs,preprintnumbers,nofootinbib]{revtex4-1}
\usepackage[pdftex]{graphicx}
\usepackage{caption,subcaption} 
\usepackage{bm}
\usepackage{amsmath}
\usepackage{relsize}
\usepackage{mathrsfs}
\usepackage{lineno}
\usepackage{amssymb}
\usepackage{accents}
\usepackage[ansinew]{inputenc}
\usepackage{lipsum}
\usepackage[colorlinks,pdfstartview=FitH]{hyperref}
\hypersetup{linkcolor=blue,citecolor=blue,filecolor=black,urlcolor=blue}
\usepackage{float}
\usepackage{setspace}
\usepackage{enumitem}

\IfFileExists{dsfont.sty}
	{\usepackage{dsfont}
         \let\mathbb=\mathds
         \newcommand{\id}{\mathds{1}}}
	{\typeout{Package dsfont.sty was not found, using alternative macros.}
         \let\mathds=\mathbb
         \newcommand{\id}{\mbox{1 \kern-.59em {\rm l}}}}
         \newcommand{\Z}{\mathds{Z}}

\renewcommand\a{\alpha}
\renewcommand\b{\beta}
\renewcommand\d{\delta}

\renewcommand\l{\lambda}
\renewcommand\r{\rho}

\renewcommand\t{\tau}
\newcommand\y{\upsilon}
\renewcommand\j{\psi}
\renewcommand\th{\theta}

\renewcommand\c{\chi}
\newcommand\ch{\raisebox{0.7\depth}{$\chi $}}
\newcommand\e{\epsilon}
\newcommand\g{\gamma}

\newcommand\m{\mu}
\newcommand\n{\nu}
\newcommand\x{\xi}
\newcommand\p{\pi}
\newcommand\h{\eta}
\newcommand\s{\sigma}

\newcommand\w{\omega}
\newcommand\ve{\varepsilon}

\newcommand\tg{\gamma_{_{D+1}}}
\newcommand\tm{\tilde\mu}
\newcommand\tn{\tilde n}
\newcommand\tA{\tilde A}
\newcommand\tJ{\tilde J}
\newcommand\tj{\tilde j}

\newcommand\tQ{\tilde Q}

\renewcommand\L{\Lambda}

\newcommand\D{\Delta}
\newcommand\G{\Gamma}
\newcommand\F{\Phi}

\newcommand\W{\Omega}

\newcommand{\lag}{\langle}
\newcommand{\rag}{\rangle}

\newcommand{\cA}{{\mathscr A}}
\newcommand{\cD}{{\cal D}}
\newcommand{\cH}{{\cal H}}
\newcommand{\cL}{{\cal L}}

\newcommand{\cT}{{\cal T}^*}

\newcommand{\bk}{{\bm k}}

\newcommand{\by}{{\bf y}}

\newcommand{\bB}{{\bm B}}
\newcommand{\bE}{{\bm E}}

\renewcommand{\vec}{\boldsymbol}
\newcommand{\vx}{\vec x}
\renewcommand{\part}{{\rm part}}

\newcommand{\be}{\begin{equation}}
\newcommand{\ee}{\end{equation}}
\newcommand{\bes}{\begin{subequations}}
\newcommand{\ees}{\end{subequations}}
\newcommand{\bea}{\begin{eqnarray}}
\newcommand{\eea}{\end{eqnarray}}

\newcommand{\pa}{\partial}
\newcommand{\inv}[1]{\frac{1}{#1}}

\newcommand{\nn}{\nonumber \\}
\newcommand{\na}{\nabla}

\newcommand{\bpsi}{\overline{\psi}}
\newcommand{\var}[2]{\frac{\d #1}{\d #2}}	
\newcommand{\sdfrac}[2]{\mbox{\small$\displaystyle\frac{#1}{#2}$}}

\newcommand{\aco}[2]{\left\{#1,#2\right\}}				

\def\nbox#1#2{\vcenter{\hrule \hbox{\vrule height#2in
\kern#1in \vrule} \hrule}}
\def\sq{\,\raise.5pt\hbox{$\nbox{.10}{.10}$}\,}
\def\sqb{\,\raise.5pt\hbox{$\overline{\nbox{.09}{.09}}$}\,}

\allowdisplaybreaks

\begin{document}

\preprint{LA-UR-19-27117}

\author{Emil Mottola$^{\rm a}$}
\email{emil@lanl.gov}
\author{Andrey V. Sadofyev$^{\rm a,b}$}
\email{sadofyev@itep.ru}
\affiliation{$^{\rm a}$Theoretical Division, T-2,  MS B283, Los Alamos National Laboratory,  Los Alamos,  NM 87545, USA}
\affiliation{$^{\rm b}$Institute for Theoretical and Experimental Physics,  Moscow,  117218,  Russia}

\title{Chiral Waves on the Fermi-Dirac Sea:\\
Quantum Superfluidity and the Axial Anomaly}
\vspace{5mm}

\begin{abstract}
\vspace{5mm}
We show that as a result of the axial anomaly, massless fermions at zero temperature define a relativistic quantum superfluid. The anomaly
pole implies the existence of a gapless Chiral Density Wave (CDW), {\it i.e.}~an axion-like acoustic mode of an irrotational and dissipationless Hamiltonian 
perfect fluid, that is a correlated fermion/anti-fermion pair excitation of the Fermi-Dirac sea. In $D\!=\!2$ dimensions the chiral superfluid effective action 
coincides with that of the Schwinger model as $e\rightarrow 0$, and the CDW acoustic mode is precisely the Schwinger boson. Since this identity holds 
also at zero chiral chemical potential, the Dirac vacuum itself may be viewed as a quantum superfluid state. The CDW collective boson is 
a $U(1)$ chiral phase field, which is gapless as a result of a novel, non-linear realization of Goldstone's theorem, extended to this case 
of symmetry breaking by an anomaly. A new local form of the axial anomaly bosonic effective action in any $D$ even spacetime is given, consistent
with superfluidity, and its quantization is shown to be required by the anomalous Schwinger terms in fermion current commutators. In QED$_4$ this 
collective Goldstone mode appears as a massless pole in the axial anomaly triangle diagram, and is responsible for the macroscopic 
non-dissipative currents of the Chiral Magnetic and Chiral Separation Effects, as well as the Anomalous Hall Effect. In a constant uniform magnetic field 
an exact dimensional reduction from $D\!=\!4$ to $D\!=\!2$ occurs and the collective $e^+e^-$ CDW chiral pair excitation propagating along the magnetic 
field direction is a Chiral Magnetic Wave, which acquires a mass gap $M^2\!=\! e^{3}B/2\pi^{2}$. Possible realizations and tests of the theory of collective 
bosonic excitations due to the anomaly in Dirac/Weyl materials are briefly discussed.

\end{abstract}

\maketitle

\tableofcontents
\newpage

\section{Introduction}

The derivation of macroscopic fluid hydrodynamics from microscopic quantum field theory is of fundamental importance, underlying many branches of physics.
This is a formidable problem in general, spanning a very large range of scales. The relationship between the macroscopic and microscopic realms becomes
more immediate at very low temperatures, approaching absolute zero, where quantum behavior dominates, and can be responsible for long range 
collective phenomena. 

An example of this closer connection at low temperatures and finite densities is provided by macroscopic superfluid behavior, characterized experimentally 
by persistent, non-dissipative currents and gapless sound modes \cite{khalatnikov2000introduction}. Theoretically these striking features are consequences 
of the spontaneous breaking of the continuous global $U(1)$ phase symmetry associated with the conservation of particle number, and the formation of a 
macroscopic coherent condensate exhibiting off-diagonal long range order \cite{Penrose51,PenroseOnsager:1956,Yang:1962,Brownbook:1992}. 
The $U(1)$ phase is a field describing an acoustic density wave excitation that is identified as a gapless Nambu-Goldstone boson 
\cite{Nambu:1961tp,Goldstone1961,Goldstone:1962es}. Although condensation to a superfluid state is a property of systems obeying Bose statistics, 
fermions such as $^3$He can form Cooper pairs that act as effective bosons which condense and also form superfluids \cite{RevModPhys.80.1215}.

Our first purpose in this paper is to show that the macroscopic properties of a relativistic superfluid are satisfied in fundamental theories such as 
quantum electrodynamics (QED), with weakly interacting massless fermions that possess an anomaly in their axial currents 
$\tilde J^{\l} = \bar \j \g^{\l}\tg \j$, when coupled to an external gauge potential $A_\l$. Here $\tg$ is the Dirac chirality matrix equal to $\g_5$ in $D\!=\!4$  
dimensions, defined in general even $D$ by eq.~(\ref{gamtilD}), and the chiral current $\tilde J^{\l}$ is the Noether current corresponding to the global 
$U^{ch}(1)$ chiral rotation $\j \rightarrow e^{i \a \tg}\, \j$ of the Dirac fermion field, which rotates left and right handed fermions by an opposite phase.  

In $D\!=\!2$ spacetime ($d\!=\!1$ space) dimensions we prove that the relationship between quantum superfluidity and massless fermion QED$_2$ is an {\it identity}. 
In fact, the effective action of the superfluid description coincides with that of the bosonized Schwinger model of massless QED$_2$~\cite{Schwinger:1962tp} in the limit 
of vanishingly small electric charge coupling $e\!\to\!0$. In other words, in this case at least, macroscopic fluid behavior is {\it directly derivable} from the microscopic theory,
and the axial anomaly is the bridge spanning scales that makes this connection possible. At non-zero chiral density $\tn$, and chemical potential $\tm$,
the axial anomaly necessarily leads to a propagating Chiral Density Wave (CDW), which is a bosonic collective mode comprised of fermion/hole pair excitation of the Fermi sea. 

The fact that $D\!=\!2$ fermions form a (Luttinger) liquid with a bosonized CDW has been anticipated \cite{Haldane:1981zza}. However, the relation to the Schwinger model, 
the essential role of its axial anomaly and precise identification with the anomalous chiral superfluid effective action has not been demonstrated previously.  
More remarkable still, since the superfluid acoustic mode depends only upon the ratio $\tm/\tn = \pi$, which is a fixed constant for free fermions in $D\!=\!2$, the superfluid 
description extends also to limiting case of $\tm = \pi \tn\rightarrow 0$, in which case the CDW becomes a fermion/anti-fermion pair excitation of the Dirac sea, implying that 
the {\it Dirac vacuum itself may also be regarded as a kind of superfluid state}. This is demonstrated in Secs.~\ref{Sec:Ideal}-\ref{Sec:ChiFlu2}. To our knowledge this is the 
first instance in which a connection between macroscopic superfluid behavior and the microscopic quantum fermion vacuum has been rigorously established.

The existence of a gapless collective boson CDW arising from the axial anomaly immediately raises a second fundamental question, namely the applicability of Goldstone's theorem.
In the more familiar case of Spontaneous Symmetry Breaking (SSB), the vacuum or low temperature ground state of the system is described by a scalar order parameter, namely the 
expectation value $\lag \F \rag$ which is non-invariant under a $U(1)$ global symmetry at the minimum of an effective potential \cite{Schmitt:2014}, but the Ward Identities are 
exactly preserved. A massless Goldstone boson then follows in dimensions $D\!>\!2$ \cite{Brownbook:1992}. This is in contrast to symmetry breaking by the fermionic axial anomaly 
-- Anomalous Symmetry Breaking (ASB) -- where the naive chiral Ward Identities are {\it explicitly violated} \cite{Adler:1969gk,treiman2015lectures,Bertlbook}, and there is no 
readily apparent effective potential to be minimized for a scalar order parameter. The standard form of Goldstone's theorem does not obviously apply to this case, and the origin of the 
gapless bosonic excitation in fermionic ASB therefore is more subtle.

That a gapless boson is implied by the anomaly was first recognized in the anomalous triangle diagram of massless QED in $D\!=\!4$ by a dispersive approach, as an {\it infrared}
singularity, independent of any ultraviolet (UV) regularization \cite{Dolgov:1971ri}. General arguments of analyticity and unitarity lead to the same conclusion 
\cite{Frishman:1980,ColemanGrossman:1982}, linking short distance to long distance behavior by the `t Hooft consistency condition \cite{Hooft:1980}, and the Adler-Bardeen 
theorem \cite{Adler:1969er}. In Refs.~\cite{Giannotti:2008cv} and \cite{Blaschke:2014ioa} it was demonstrated that the massless boson pole of the axial anomaly in both $D\!=\!2$ 
and $D\!=\!4$ dimensions arises from correlated pairs of massless fermions and anti-fermions (particle-hole pairs in the condensed matter context) traveling together co-linearly
at the speed of light. In other words, the axial anomaly itself implies fermion pairing and the existence of a propagating gapless effective boson excitation, analogous to Cooper 
pairing, even in the limit of very weak (or in the limiting case of absence of) fermion self-interactions. This gapless excitation is made explicit in $D\!=\!2$ by the 
techniques of bosonization \cite{Coleman:1974bu}. In $D\!=\!4$ a bosonic form of the effective action of the axial anomaly was given in  \cite{Giannotti:2008cv}.
It has become recognized only relatively recently that the appearance of a collective massless boson mode, not apparent in the classical Lagrangian, is a general feature 
of both axial and conformal anomalies, whose effects can extend to long distance or macroscopic scales \cite{Mottola:2006ew,ArmCorDRosaPRD10,Mottola:2010gp}. 
Since the distinguishing feature of SSB, a gapless excitation, is present in massless fermion theories with an axial anomaly, it suggests that Goldstone's theorem 
can be generalized and extended to anomalous fermion theories, despite there being no effective potential or scalar order parameter immediately apparent at the outset. 

Demonstrating that Goldstone's theorem can indeed be generalized to ASB is our second main purpose in this paper. The novel extension of Goldstone's theorem 
in the case of ASB differs from SSB in that it rests upon the {\it anomalous} current commutators, or Schwinger terms, in the underlying fermion theory, which remain 
non-vanishing even in the weak coupling limit $e\!\to\!0$ and vanishing background gauge field, where the divergence $\pa_\l \tilde J^{\l} = \bar \j \g^{\l}\tg \j \to 0$.  
In this limit there is a propagating (pseudo-)scalar CDW, and this bosonic collective mode is associated with fermion condensation and $\lag \bar\j \j\rag \neq 0$.
This extension of Goldstone's Theorem to ASB is proven in Sec.~\ref{Sec:NambGold}.

The weak coupling limit $e\!\rightarrow\!0$ is subtle, particularly in $D\!=\!2$ dimensions, where the Mermin-Wagner-Coleman (MWC) theorem \cite{Mermin1966,Coleman:1973}
tells us that there can be no true long-range order and no true Goldstone bosons. The $D\!=\!2$ case therefore merits special attention. Since the would-be Goldstone mode is
just the boson of the Schwinger model in $D\!=\!2$, it becomes massive or gapped with $M^2 = e^2/\pi$, so that for any finite $e$, however small, the Goldstone mode becomes 
`Higgsed,' there is no long range order and no gapless mode at distances greater than $1/M$, consistent with the MWC theorem. Nevertheless as $e\!\to\!0$ but $L\to\infty$ 
with $eL\gtrsim 1$ fixed, the CDW approximates a Goldstone mode over a larger and larger range of distance scales with only algebraic power law decay of the correlator. 
This is the behavior known as `quasi-long-range order' \cite{SachdevBook:2011}. 

The third major goal of this paper is to extend the derivation of macroscopic behavior from quantum field theory to higher (even) dimensions larger than two, again 
relying upon the axial anomaly and the anomaly pole. We show in Sec.~\ref{Sec:ChiAnom4} that bosonization and macroscopic superfluidity effects persist in higher dimensions 
as well, as a direct result of the axial anomaly, at least at $T\!=\!0$ for massless fermions, in a subsector of the theory at weak enough coupling that self-interactions can be neglected. 
This relies upon a new form of the effective bosonic action of the axial anomaly in any $D\!=2n$ even dimensions in the longitudinal sector of the axial current, where the massless 
boson anomaly pole of the triangle $\lag \tilde J^\l J^\a J^\b\rag$ diagram resides. This collective boson in $D\!=\!4$, again composed of massless fermion pairs, is entirely responsible 
for the non-dissipative, macroscopic Chiral Magnetic and Chiral Separation currents of QED$_4$ \cite{Kharzeev:2015znc, Huang:2015oca}. That the same $D\!=\!4$ effective 
action and massless pole of the triangle anomaly of QED$_4$ is also responsible for the Anomalous Hall Effect in quantum materials such as Weyl semi-metals 
\cite{Landsteiner:2016led} is shown in Sec.~\ref{Sec:AHE}, in another example of a macroscopic quantum effect in $D>2$ directly traceable to the microscopic anomaly.

In the case of a constant and uniform magnetic field $\vec B$ background with a parallel $\vec E$ field independent of the transverse coordinates, the $D\!=\!4$ triangle anomaly 
reduces to the $D\!=\!2$ case. The CDW along the common $\vec E\parallel \vec B$ direction is a Chiral Magnetic Wave (CMW), coinciding with the Schwinger boson collective 
excitation in $D\!=\!2$; or in other words this is an example of {\it dimensional reduction}, where macroscopically observable quantum coherence effects 
resulting from the microscopic QFT anomaly in $D>2$ dimensions can be rigorously related back to the $D\!=\!2$ case. The role of the dimensionful $D\!=\!2$ coupling
of the Schwinger model is taken by $2\a eB$. This is also a new result, rigorously proven for the first time to our knowledge in Sec.~\ref{Sec:DimRed}.

As a final example of macroscopic effects in $D\!=\!4$, we consider in Sec.~\ref{Sec:ChiFlu4} free massless fermions at finite chiral density in QED$_4$, where the collective boson 
predicted by the Goldstone theorem extended to ASB of Sec.~\ref{Sec:NambGold} is a gapless acoustic mode propagating at the sound speed $v_s^2 = \frac{dp}{d\ve} =\frac{1}{3}$,
sourced by and coupled to $\vec E \cdot \vec B$. Possible realizations of the CDW as an axion-like mode in Weyl materials are briefly discussed in Sec. \ref{Sec:Sum},
in the Summary and Outlook. 

There are three Appendices containing some additional technical details, included for completeness and supplementing the main text. Appendix A 
clarifies the relationship between the superfluid energy-momentum tensor and Hamiltonian. Appendix B collects a number of useful formulae for free fermions in two dimensions 
and their bosonization for reference. Appendix C examines the delicate limit of $e \to 0$ in a finite volume system in $D\!=\!2$, the $\th$-vacuum periodicity arising from 
topologically non-trivial gauge transformations, and fate of the Nambu-Goldstone mode in the Schwinger model in the infinite spatial volume limit.

\vspace{3mm}
\noindent
{\it Note on Organization of the Paper:}

In the interest of making the paper as self-contained and accessible to as wide a cross-disciplinary audience as possible, review of some topics, such as the Schwinger 
model and triangle anomaly seemed necessary, particularly since they may not be familiar to readers with different backgrounds. For the most part this supplementary 
review material is relegated to the Appendices. A Table of Contents and brief account of the main new results and their importance are provided both in this Introduction, 
and summarized again in Sec.~\ref{Sec:Sum}, with references to the specific numbered relations establishing these results in the text, for the convenience of readers 
who may wish to skip directly to those sections containing the results in which they are most interested.

\vspace{-2mm}
\section{Ideal Hydrodynamics of An Anomalous Chiral Fluid}
\label{Sec:Ideal}
\vspace{-3mm}
\subsection{Euler-Lagrange Action Principle}
\vspace{-2mm}

We begin by giving the action principle that provides a consistent framework for relativistic ideal chiral fluids incorporating the axial anomaly. This is a 
new gauge invariant formulation which unlike some earlier treatments, {\it e.g.} \cite{Monteiro:2014wsa}, conserves electric charge. It extends and 
generalizes earlier discussions of relativistic but non-anomalous or non-chiral fluids \cite{Schutz:1970my,Jackiw:2004nm}. 
Those were based on generalizing to Lorentz invariant systems the Euler-Lagrange action principle for non-relativistic fluids \cite{Clebsch}. The key ingredient
of these variational principles is the introduction of a `dynamic velocity' vector field $\vec \x$ that can be expressed in terms of scalar (Clebsch) potentials
\cite{Seliger1968}. In its relativistic generalization the dynamic velocity $\x_\l$ is a covariant $D$-vector in $D$ spacetime dimensions that couples 
to the conserved current whose hydrodynamic flow is under study. In this paper the fermionic axial current $\tilde J^\l= \bar\psi \g^\l\tg \psi$ is our primary focus. 

In general several Clebsch potentials, denoted by $\h,\a,\b,\dots$ are necessary to describe arbitrary rotational motions of a fluid and entropy flow at finite 
temperatures, in terms of which the dynamic velocity can be expressed as $\xi_\l = -\pa_\l\h + \a\,\pa_\l \b + \dots$\cite{Schutz:1970my}. 
However the essential feature of a superfluid is that it should be {\it dissipationless}. This implies that the $\a,\b, \dots$ terms describing entropy or rotational currents
are negligible, and should vanish entirely at zero temperature. In that case the dynamic velocity field can be expressed as a pure gradient of a single scalar potential 
to be denoted here by $\h$, so that we may take $\x_\l = -\pa_\l \h$. Then the minimal action for ideal hydrodynamics of a chiral superfluid in $D=d+1$ (even) 
spacetime dimensions is
\be
S_{\c\rm fl}= \int\!\! d^D\!x\, \cL_{\c\rm fl} = \int\!\! d^D\!x \left\{ \big(\pa_\l \h+ \tA_\l\big)\tilde J^\l+ \h\,{\cA}_D \!-\! \ve (\tn)\right\}
\label{Sfluid}
\ee
where the tildes denote chiral quantities, $\tn$ is the chiral number density and $\ve (\tn)$ is the equilibrium energy density of the fluid in its rest frame.
An external axial potential $\tA_\l$ is introduced so that
\be 
\frac{\d S_{\c\rm sf}}{\d \tA_\l} = \tJ^\l = \tn\,  u^\l
\label{J5n5}
\ee
is defined, with $u^\l$ the relativistic kinetic velocity of the fluid. Since the relativistic velocity is normalized by $u^\l u_\l = u^\l  g_{\l\n} u^\n= -1$ 
in units in which the speed of light $c=1$, with $g_{\l\n}$ the spacetime metric, the chiral number density can be expressed in the relativistically invariant form
\vspace{-2mm}
\be
\tn \equiv \left(-\tJ^\l \tJ_\l\right)^{\frac{1}{2}}
\label{n5def} 
\ee
in terms of the axial current. The ${\cA}_D$ term takes account of the axial current anomaly in (\ref{anomJ5}) below.
For some related discussions of effective actions for anomalous fluids, see \cite{Lublinsky:2009wr, Dubovsky:2011sk, Monteiro:2014wsa, Glorioso:2017lcn}.

The variational principle for the fluid action (\ref{Sfluid}) requires that the pseudoscalar Clebsch potential $\h$ and the axial current $\tJ^\l$ 
be varied independently, in addition to the variation (\ref{J5n5}). Variation of (\ref{Sfluid}) with respect to $\h$ gives 
\vspace{-1mm}
\be
\pa_\l \tJ^\l = \cA_D \stackrel{D=2n}{=} \frac{2}{(4\p)^n\, n!}\,\e^{\m_1\n_1...\m_n\n_n}F_{\m_1\n_1}...F_{\m_n\n_n}
\label{anomJ5}
\ee
where $\cA_D$ is the axial anomaly for massless Dirac fermions, $F_{\m\n} = \pa_\m A_\n - \pa_\n A_\m$ is the electromagnetic field strength,
and $\e^{\m_1\n_1...\m_n\n_n}$ is the totally anti-symmetric Levi-Civita tensor in $D\!=\!2n$ even dimensions \cite{Zumino:1983rz}.
We focus on the $D=2, 4$ cases explicitly in this paper, providing a basis for the superfluid action (\ref{Sfluid}) 
from QFT first principles, by which $\h$ will be identified with the dynamical phase field of chiral symmetry breaking. 

Variation of (\ref{Sfluid}) with respect to the axial current $\tJ^\l$ gives
\be
\frac{\d}{\d \tJ^\l} S_{\c\rm fl} =  \pa_\l\h  +  \tA_\l - \left(\frac{d\ve}{\,d \tn}\right)\left( \frac{d\tn}{d\tJ^\l}\right)
=  \pa_\l\h  +  \tA_\l + \sdfrac{\tm}{\tn} \, \tJ_\l=0
\label{varJ}
\ee
in view of (\ref{n5def}), where
\vspace{-8mm}
\be
\tm \equiv \frac{d\ve}{d \tn}
\label{mu5def}
\ee
is the equilibrium chiral chemical potential. Thus using (\ref{J5n5}), the dynamic velocity field is
\be
\x_\l = -\pa_\l \h = \tm\, u_\l 
\label{dynvel} 
\ee
where we now set the external axial potential $\tA_\l = 0$ here and in the following. We also have
\be
\tm = \big(\!-\x_\l\x^\l\big)^{\frac{1}{2}} = \left(-\pa^\l\h\,\pa_\l\h\right)^{\frac{1}{2}}
\label{mu5eta}
\ee
so that the chiral chemical potential is also a relativistic invariant, analogous to (\ref{n5def}). 

If (\ref{Sfluid}) is evaluated at the extremum (\ref{varJ}), denoted by an overline 
\be
\overline S_{\c\rm fl} \Big\vert_{\cA_D=0}= \int\! p \, d^D\! x= \int \!dt \!\int\! p \,d^d \vx = - \int \!dt\ \W
\label{Sp}
\ee
where 
\vspace{-8mm}
\be
p(\tm)= \tm\,\tn - \ve = \tn^2\, \sdfrac{d }{d \tn} \Big(\frac{\ve}{\tn}\Big) \qquad {\rm satisfying}\qquad \frac{dp\,}{d\tm} = \tn
\label{pres}
\ee
is the pressure of the fluid when the external field anomaly term $\cA_D$ is set to zero, and $\W$ is its Grand Potential.
In equilibrium at finite temperature $\int \! dt$ is replaced by $\b = \hbar/k_BT$ by continuation to imaginary time.
However, in this paper we work in real time and at zero temperature. 

\vspace{-5mm}
\subsection{Canonical Hamiltonian}
\label{Sec:CanHam}
\vspace{-2mm}

The dissipationless nature of the perfect chiral fluid described by (\ref{Sfluid}) is made explicit by its Hamiltonian form 
\cite{Schutz:1971ac, Jackiw:2004nm, Monteiro:2014wsa}. Defining the momentum conjugate to $\h$
\vspace{-2mm}
\be
\Pi_{\h} \equiv \frac{\d }{\d \dot \h}\,S_{\c\rm fl} = \tJ^0
\label{J05}
\ee
we find the Hamiltonian density
\vspace{-3mm}
\be
\cH_{\c\rm fl} =  \Pi_{\h} \,\dot \h - \cL_{\c\rm fl} =  - \,\tJ^i\, \na_i \h  - \h\, \cA _D+  \ve  
\label{Hamanom}
\ee
at $\tA_\l=0$. To express this in terms of the canonical pair $(\h, \Pi_\h)$ we first solve (\ref{varJ}) for 
\vspace{-2mm}
\be
\tJ^i = -\frac{\tn}{\tm}\, \na^i \h
\label{J5i}
\ee
again at $\tA_\l=0$, so that from (\ref{n5def}) and (\ref{J05}) we have
\be
\Pi_\h ^2 = \tn^2 + \tJ_i\tJ^i = \tn^2 + \frac{\tn^2}{\tm^2}\, (\vec\na \h)^2 =  \left(\sdfrac{\tn}{\tm} \right)^2 \big[ \tm^2 + (\vec\na \h)^2 \big]
\ee
from which
\vspace{-8mm}
\be
\frac{\tn}{\tm}= \frac{1}{\tm}  \frac{dp}{d\tm}= \frac{|\Pi_\h |}{\sqrt{\tm^2 + (\vec\na \h)^2}}\ge 0
\label{n5m5}
\ee
where the positive square root is always taken. Thus from (\ref{pres}) we obtain
\be
\ve - \,\tJ^i\, \na_i \h = \tm\,\tn - p + \sdfrac{\tn}{\tm}\,  (\na^i \h) \,(\na_i \h) = \sdfrac{\tn}{\tm}\, \big[ \tm^2 + (\vec\na \h)^2 \big] - p
\label{m5n5J}
\ee
and making use of this and (\ref{n5m5}), (\ref{Hamanom}) becomes
\be
\cH_{\c\rm fl}[\h, \Pi_\h] = \vert \Pi_\h \vert \sqrt{\tm^2 +  (\vec\na \h)^2}\,- p(\tm) - \h\, \cA _D
\label{Hamcsf}
\ee
where $\tm[\na\h, \Pi_\h]$ and $p(\tm)$ are to be regarded here as implicit functions of $\Pi_\h$ and the spatial gradient $\vec \na\h$
through (\ref{n5m5}), once the equilibrium functional form of $p=p(\tm)$ is specified.

The fluid Hamiltonian is $H_{\c \rm fl} \!= \!\int\!d^d {\vec x}\,  \cH_{\c \rm fl}$ from which Hamilton's eqs.~follow, namely
\bes
\bea
\dot \h = \frac{\d }{\d \Pi_\h}H_{\c\rm fl} = \frac{\tm}{\tn}\, \Pi_{\h} &=& {\rm sgn}(\Pi_\eta)\sqrt{\tm^2 +  (\na \h)^2}\label{hdot} \\
\dot \Pi_\h = -\frac{\d }{\d \h}H_{\c\rm fl}  &=& \vec\na \!\cdot\! \left(\frac{\tn}{\tm}\, \vec\na\h\right) + \cA_D\label{rh5dot}
\eea
\label{eom}\ees
for the canonical pair $(\h, \Pi_\h)$, where we have used the fact that the variation of the $\tm$ dependence drops 
out upon using (\ref{n5m5}). Eq.~(\ref{hdot}) recovers the time component of (\ref{varJ}), so that~(\ref{rh5dot}) is  
\be
\frac{\pa}{\pa t} \left(\frac{\tn}{\tm}\, \frac{\pa \h}{\pa t}\right) - \vec\na \!\cdot\! \left(\frac{\tn}{\tm}\, \vec\na\h\right) = \pa_\l \tJ^\l =\cA_D
\label{etawave}
\ee
which recovers the axial current anomaly (\ref{anomJ5}), upon making use of (\ref{J05}) and (\ref{J5i}). 

When applied to long wavelength small perturbations $\d\h$ away from equilibrium -- the limit in which the fluid description should be valid -- (\ref{etawave}) 
describes a gapless CDW acoustical mode with $\cA_D$ as its source. Thus the perfect chiral fluid hydrodynamics determined 
by (\ref{Sfluid}) is both irrotational and dissipationless, with a time reversible Hamiltonian dynamics (\ref{eom}) and a gapless excitation. 
These are necessary features of a relativistic chiral superfluid \cite{Alford:2012vn, Schmitt:2014}. In Sec.~\ref{Sec:NambGold} we show that $\h$ 
is a phase field associated with spontaneous breaking of $U^{ch}(1)$ symmetry, giving rise to a Nambu-Goldstone mode, also as expected
for superfluidity.

The Hamiltonian fluid acoustic mode is quantized by replacing the Poisson bracket of the canonical pair $\{\h, \Pi_\h\}$ by
their equal time commutator
\vspace{-2mm}
\be
\big[\h(t, \vx), \Pi_\h (t, \vx')\big] = \big[\h(t, \vx), \tJ^0 (t, \vx')\big] = i\,\d^d (\vx - \vx') 
\label{etaPi}
\vspace{-2mm}
\ee
(in units where $\hbar =1$). We show in Secs.~\ref{Sec:Schw} and \ref{Sec:ChiAnom4} that this canonical commutator of the 
bosonic hydrodynamic description is in fact {\it required} by the anomalous current commutators of the underlying fermionic
description of a {\it relativistic quantum anomalous chiral superfluid}.

\vspace{-4mm}
\subsection{Chiral Superfluid Energy-Momentum Tensor}
\label{App:EMT}
\vspace{-2mm}

The energy-momentum tensor for the chiral fluid may be found by generalizing the effective hydrodynamic action (\ref{Sfluid})
to curved spacetime by the Equivalence Principle, given by \cite{Schutz:1970my,Jackiw:2004nm, Monteiro:2014wsa}
\be
S_{\c\rm sf}= \int\! d^D\!x \sqrt{-g} \left\{\tJ^\l\,\pa_\l\h - \ve (\tn)\right\} + \int\!d^D\!x\, \h\,\cA _D
\ee
where we have set $\tA_\l=0$, and used the fact that the axial anomaly term $\cA _D$ in (\ref{anomJ5}) is directly a tensor density 
and thus does not acquire a $\sqrt{-g}\equiv \sqrt{- {\rm det}\,(g_{\l\n})}$ metric factor. It follows that
\be
T^{\l\n} = \frac{2\!\!}{\sqrt{-g}}\frac{\d S_{\c\rm sf}}{\d g_{\l\n}} = p\,g^{\l\n} + (p+\ve) \,u^\l u^\n= p\,g^{\l\n} + \frac{\tn}{\tm}\, \pa^\l\h\, \pa^\n\h
\label{EMT}
\ee
where in performing the metric variation in (\ref{EMT}) $\pa_\l\h$ and $\tJ^\l$ are taken to be metric independent, while $\tJ_\l = g_{\l\n}\tJ^\n$. 
The anomaly term $\cA_D$, lacking a $\sqrt{-g}$ factor, does not contribute to $T^{\l\n}$ in the variation (\ref{EMT}), but does give
rise to the electromagnetic current 
\be
J^\n = \frac{\d S_{\c\rm sf}}{\d A_\n} = \frac{\d }{\d A_\n}\int d^D\! x\, \h\, \cA _D
\label{Janom}
\ee
dependent upon $\h$. The energy-momentum tensor (\ref{EMT}) is that of a perfect chiral fluid.

The divergence of (\ref{EMT}) evaluated in flat spacetime is
\be
\pa_\n T^{\l\n} = \pa^\l p + \pa_\n\left(\frac{\tn}{\tm}\ \pa^\n\h\right)\pa^\l\h+\frac{\tn}{\tm}\,(\pa^\n\h) (\pa_\l\pa_\n\h)\,.
\label{Tdiv}
\ee
Using (\ref{mu5eta}) and (\ref{pres}), the pressure gradient in this expression is
\be
\pa^\l p= \frac{dp\,}{d\tm\!}\ \pa^\l\tm = -\frac{\tn}{\tm}\, \big(\pa^\n\h\big) \big( \pa_\l\pa_\n\h\big)
\ee
which cancels the last term in (\ref{Tdiv}), resulting in
\be
\pa_\n T^{\l\n}=\pa_\n\left(\frac{\tn}{\tm}\, \pa^\n\h\right)\pa^\l\h = -\big(\pa_\n \tJ^\n\big)\big(\pa^\l\h\big) = - \cA_D\, \pa^\l\h
\label{nconsA}
\ee
where the relation (\ref{varJ}) between $\tJ^\n$ and $\pa^\n\h$ and the anomalous divergence (\ref{anomJ5}) have been used.
Eq.~(\ref{nconsA}) together with the anomaly eq.~(\ref{anomJ5}) show that if $\cA_D$ vanishes, so that $\tJ^\n$ is conserved,
then the energy-momentum tensor $T^{\l\n}$ is conserved as well. In the presence of an external, {\it i.e.} non-dynamical, 
electromagnetic field the energy-momentum tensor (\ref{nconsA}) should satisfy
\vspace{-3mm}
\be
\pa_\n T^{\l\n}= F^{\l\n}J_{\n}
\label{nconsJ}
\vspace{-4mm}
\ee
where $J_\n$ is the electromagnetic current (\ref{Janom}) induced by the anomaly. Proof of the equality of (\ref{nconsA}) and (\ref{nconsJ}),
as well as the reason for the difference between the Hamiltonian (\ref{Hamcsf}) and $T^{00}$ are given in Appendix \ref{Sec:Schouten},
as both depend upon the special properties of the axial anomaly $\cA_D$.

\vspace{-3mm}
\section{Quantum Anomalous Chiral Superfluid in Two Dimensions}
\label{Sec:ChiFlu2}
\vspace{-2mm}

In this section we show that the bosonized form of massless free fermions in $D\!=\!2$ in fact {\it coincides} with the effective chiral fluid
description of the previous section, thus deriving it completely from first principles of microscopic fermion QFT in the $D\!=\!2$ case.

\vspace{-4mm}
\subsection{The Schwinger Model and its Axial Anomaly}
\label{Sec:Schw}
\vspace{-4mm}

Electrodynamics in $1+1$ dimensions, QED$_2$, is defined by the classical action
\vspace{-3mm}
\be
S_{cl} = \int\!d^2x \, \bpsi\,\Big\{i\g^a \big(\!\stackrel{\!\!\leftrightarrow}{\pa_a}\!\!-i A_a\big) - m\Big\}\psi - \frac{1\,}{4e^2}\! \int\!d^2x\, F_{ab}F^{ab}
\label{Schwmod}
\vspace{-3mm}
\ee
and is exactly soluble for vanishing fermion mass $m=0$ \cite{Schwinger:1962tp}. The solution relies upon the special property of the Dirac matrices
$\g^{a}\tg = -\e^{ab}\g_b$ where $\tg = \g^0\g^1$ in $D\!=\!2$ dimensions.\footnote{Notation: Indices $a,b =0,1;$ Metric $g_{ab}  = {\rm diag} (-1,1)\!=\! g^{ab}; 
\,\e_{ab}= - \e_{ab};\,\e_{01} =  +1= \!-\e^{01}; \,F_{ab} \equiv \pa_a A_b - \pa_b A_a; \\
\stackrel{\!\!\leftrightarrow}{\pa_a} \equiv \frac{1}{2} \big(\!\stackrel{\!\!\rightarrow}{\pa_a} - \stackrel{\!\!\leftarrow}{\pa_a}\!\!\big);$ 
$2 \times 2$ Dirac matrices $\g^0\! =\!\s_1\!=\!(\g^0)^\dag,  \g^1 \! = -\! i\s_2, \tg \! =\!\g^0\g^1\!=\! \s_3; $ with $\s_i, i=1, 2, 3$ the usual Pauli matrices, 
$\bpsi \equiv \psi^{\dag}\g^0$, and anti-symmetrization over $\bpsi , \psi$ is understood.  We use lower case $j^a, \tj^b$ for currents and $a, b$ for
Latin indices in $D=2$ to distinguish them from $D\ge 4$ currents $J^\n, \tJ^\l$ with Greek indices $\n,\l$, {\it etc.}} 
This property has the consequence that the chiral current 
\vspace{-2mm}
\be
\tj^a \equiv \bpsi \g^a \tg \psi = -\e^{ab} \bpsi \g_b \psi \ = -\e^{ab} j_b
\label{J5dual}
\vspace{-2mm}
\ee
is dual to the charge current $j^a\equiv \bpsi \g^a \psi$. Both currents $j^a,\tj^a$ would appear to be conserved Noether currents, corresponding 
to the classical $U(1)\otimes U^{ch}(1)$ symmetry of the Dirac Lagrangian in (\ref{Schwmod}) when $m=0$. However, both $U(1)$ 
symmetries cannot be maintained at the quantum level, and at least one of these symmetries is broken by an anomaly \cite{Johnson:1963vz}.

\begin{figure}[ht]
\vspace{-1.5cm}
\centering
\begin{subfigure}{.45\textwidth}
\centering
\includegraphics[width=.9\textwidth]{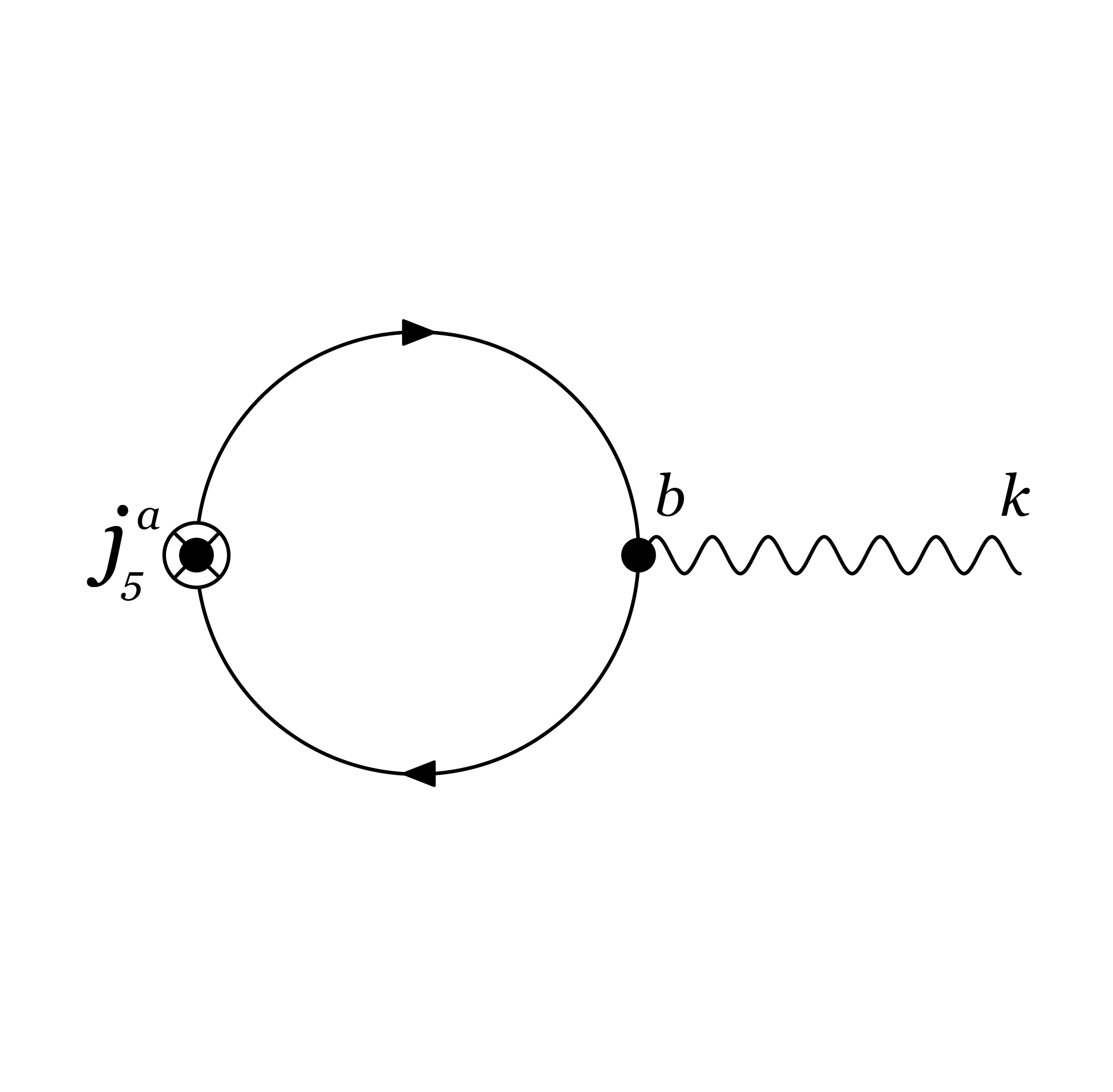}
\label{fig:JJ-loop}
\end{subfigure}
\hspace*{1cm}
\begin{subfigure}{.45\textwidth}
\centering
\includegraphics[width=.9\textwidth]{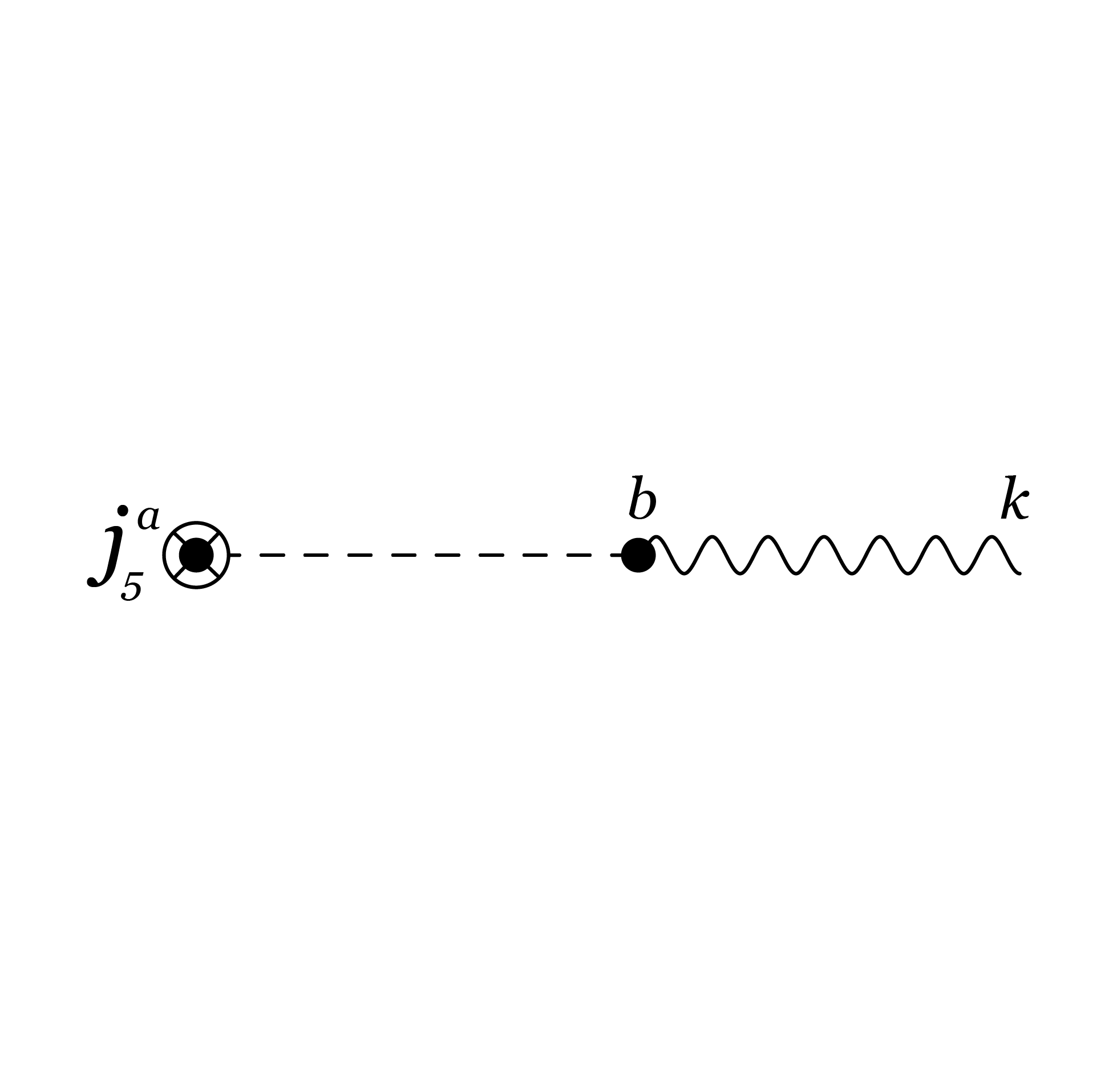}
\label{fig:JJ-tree}
\end{subfigure}
\vspace{-1.5cm}
\caption{One-loop fermion polarization diagram (a) and equivalent pseudoscalar tree diagram (b).}
\label{fig:JJ}
\end{figure}

The addition of the $F_{ab}F^{ab}$ term in (\ref{Schwmod}) and consistency of Maxwell's eqs.~$\pa_a F^{ba} = e^2 j^b$ 
requires $\pa_a j^a =0$, and the breaking of the $U^{ch}(1)$ chiral symmetry. Explicitly, if conservation of $j^a$ is enforced, 
then (\ref{J5dual}) and the one-loop vacuum polarization `diangle' diagram of Fig.~\ref{fig:JJ} implies
\bea
&&\tj^a(x)\Big\vert_{m=0} =-i  \e^a_{\ b}\! \int d^2 y\,\lag\cT j^b (x) j^c(y)\rag_{_0} \,A_c(y) \nn
&&= -\e^a_{\ b} \! \int \!d^2 y\int \!\frac{d^2k}{(2\pi)^2}\,\Pi_{2}^{bc}(k)\, e^{ik\cdot (x-y)} A_c(y)
\label{Pitil2}
\eea
where $\cT$ denotes covariant time ordering, and
\vspace{-2mm}
\be
\Pi_{2}^{bc}(k)=\frac{1}{\pi k^2}\left(k^b k^c-g^{bc}k^2\right)   
\label{Pi2}
\vspace{-2mm}
\ee
is the polarization tensor for massless fermions in $D\!=\!2$. Thus
\be
\tj^a(x)\Big\vert_{m=0} = \sdfrac{1}{\pi}\e^{ab}\!\int \!d^2y\, \sq^{-1}_{xy}\,\pa_c F^{c}_{\,\ b}(y) = \sdfrac{1}{\pi}\,\pa^{a}_x\!\! \int\!d^2y\, \sq^{-1}_{xy} \, ^*\!F(y)
\label{J5J}
\ee
where $^*\!F \equiv \frac{1}{2}\e^{ab}F_{ab}$ is the pseudoscalar dual to $F_{ab}$, and $\sq^{-1}_{xy}= -\frac{1}{4 \pi} \ln (x-y)^2 + const.$ 
is the massless scalar propagator in $D\!=\!2$. Thus if the background electric gauge field $F_{ab} \neq 0$, the chiral current (\ref{J5J}) acquires 
the finite anomalous divergence
\vspace{-2mm}
\be
\pa_a \tj^a  = \frac{1}{2\pi} \, \e^{ab}F_{ab} = \frac{^*\!F}{\pi} =  \cA_2
\label{anom2}
\vspace{-2mm}
\ee
at the one-loop level. This turns out to be an exact result in massless QED$_2$, because a further special property of massless QED$_2$ 
is that the current induced by a background gauge potential is strictly linear in $A_a$ for arbitrary $A_a$. 
Note also that $^*\!F = F_{10} = E$ is just the electric field in $d\!=\!1$ spatial dimension. 

The chiral anomaly (\ref{anom2}) corresponds to the {\it exact} non-local 1PI effective action 
\vspace{-2mm}
\be
S^{NL}[A]  = -\frac{1}{2\pi}\int\!d^2x\!\int \!d^2y\, 
{^*\!F}(x)\sq^{-1}_{xy}\,{^*\!F}(y)
\label{SanomNL}
\vspace{-2mm}
\ee
obtained by integrating out the massless fermions in the functional integral for (\ref{Schwmod}).
The appearance of the $1/k^2$ massless pole in (\ref{J5J}) corresponding to the massless scalar propagator
$ (\sq^{-1})_{xy}$ in the 1PI non-local effective action (\ref{SanomNL}) signals that an effective scalar boson degree
of freedom is associated with the anomaly. Indeed a (pseudo-) scalar boson field $\c$ may be introduced to rewrite
the non-local action (\ref{SanomNL}) in the local form \cite{Blaschke:2014ioa}
\be
S_{\rm eff}[\c; A] = 
\int\!d^2x\left\{- \sdfrac{\pi}{2}\, \big(\pa_a \c\big) \big(\pa^a\c\big) -\, ^*\!F\, \c \right\} 
\label{Seffchi}
\ee
whose variation with respect to $\c$ gives
\vspace{-5mm}
\be
\sq\c= \sdfrac{1}{\pi}\,^*\!F\,.
\label{chieom}
\vspace{-2mm}
\ee
Solving this eq.~for $\c$ by inverting $\sq$, and substituting the result back into (\ref{Seffchi}) returns (\ref{SanomNL}).
The fermion loop of the original theory (\ref{Schwmod}) is thus completely equivalent to the $\c$ boson propagator
of the effective action (\ref{Seffchi}). This equivalence is illustrated in Fig. \ref{fig:JJ}.

The axial and vector currents may be expressed in terms of the effective boson $\c$ field via
\be
\tj^a =\pa^a\c\,,\qquad\qquad j^a= \var{}{A_a}\,S_{\rm eff}=-\e^{ab}\pa_b\c 
\label{vecaxvec}
\ee 
so that the conservation of $j^a$ becomes a topological identity and the chiral anomaly (\ref{anom2}) follows from the eq.~of motion for $\c$
(\ref{chieom}). This is the chiral bosonization of fermion currents in $1+1$ dimensions \cite{Coleman:1974bu,Mandelstam:1975hb}. 
The chiral boson $\c$ obeys a massless wave eq.~(\ref{chieom}) with $\cA_2$ as its source. It acquires a mass 
$M=e/\!\sqrt{\pi}$ only if the coupling $e \neq 0$ in the full Schwinger model of (\ref{Schwmod}).

The explicit Fock space operator representation of the chiral boson in terms of fermion bilinears is given in Appendix \ref{Sec:Massless}.
These lead to the equal time commutator
\vspace{-2mm}
\be
\big[\c(t,x), \dot\c(t,x')\big] = \frac{i}{\pi} \,\d (x-x')
\label{comchi}
\vspace{-2mm}
\ee
which is exactly the canonical commutation relation obtained directly from the effective action (\ref{Seffchi}). 
From (\ref{jpsipm})-(\ref{zeromodes}) the fermion current densities in the Fock space representation are
\vspace{-2mm}
\bes
\begin{align}
j^0 &= \,\pa_x \c \, =\, \tj^1 \label{j0chi}\\
j^1 &=\!  -\pa_t \c = \tj^0
\label{j1j50}
\vspace{-2mm}
\end{align}
\label{j0j1phi}\ees
reproducing the bosonization relations (\ref{vecaxvec}) obtained in the effective action representation.  These relations and the commutator (\ref{comchi}) imply
\be
\big[ j^0 (t,x), j^1 (t,x')\big]= -\frac{i}{\pi}\, \pa_x\,\d(x-x')
\label{j0j1com}
\ee
which is the Schwinger term in the equal time commutator of the current components \cite{Schwinger:1959xd,Jo:1985}. The normal ordering prescription 
on the currents and precise definition of the fermion vacuum is essential to derive these commutation relations in the fermion representation.

Since the~{\it anomalous} Schwinger current commutator in QED$_2$ is equivalent to the {\it canonical} commutator of the chiral boson $\c$, it
must be treated as a genuinely propagating collective degree of freedom, not apparent in the original fermionic action (\ref{Schwmod}). Composed of fermion 
bilinears, analogous to Cooper pairs in condensed matter systems, $\c$ has the massless propagator (\ref{J5J}) which is associated with the axial 
anomaly \cite{Blaschke:2014ioa}. The fermion bilinears are correlated/entangled fermion pairs in the two-fermion intermediate states of the polarization 
diagram Fig.~\ref{fig:JJ}, moving co-linearly at the speed $c=1$, and equivalent to a massless boson according to (\ref{jpsipm})-(\ref{phiposn}). 
Because of (\ref{vecaxvec}) the propagating wave solutions of (\ref{chieom}) are jointly Chiral Density Waves and Charge Density Waves, arising from 
massless particle-hole excitations of the Dirac sea. As we shall now show, these same CDWs may also be regarded as acoustic modes 
in the superfluid hydrodynamic description at $T=0$, and thus the superfluid hydrodynamic description of Sec.~\ref{Sec:Ideal} applies both to the 
filled Fermi-Dirac sea at non-zero chiral density, and the Dirac fermion vacuum itself.

\vspace{-3mm}
\subsection{Equivalence to Chiral Superfluid Hydrodynamics in $D=2$}
\label{Sec:equiv}
\vspace{-2mm}

At finite chiral chemical potential $\tm$, but $T\!=\!0$, massless fermions in $D\!=\!2$ have the energy density and pressure
({\it cf.}~Appendix \ref{Sec:Massless})
\vspace{-4mm}
\be
\ve = \frac{\pi}{2}\, \tn^2= p= \frac{1}{2 \pi}\, \tm^2   \qquad {\rm where} \qquad \tm = \pi\, \tn\,,
\label{ep2}
\vspace{-2mm}
\ee
so that (\ref{varJ}) gives at $\tA_\l =0$
\vspace{-6mm}
\be
\pa_a \h =  - \sdfrac{\tm}{\tn}\ \tj_a = -\pi\, \tj_a =- \pi\, \pa_a \c
\label{etachi}
\vspace{-4mm}
\ee
since $ \tj^a = \pa^\a\c$ by the bosonization formulae (\ref{vecaxvec}). 

The simple proportionality in (\ref{etachi}) between the gradient of the $\h$ potential and that of the {\it a priori} independent chiral boson 
field $\c$ is what makes possible the identification of the Schwinger model with the chiral superfluid of Sec.~\ref{Sec:Ideal}. 
Recalling (\ref{n5def}) with (\ref{etachi}) we now obtain
\be
(\pa_a\h)\, \tj^a- \ve (\tn) = - \pi\, (\pa_a\ch)(\pa^a\ch) + \frac{\pi}{2} \,\tj^a \,\tj_a= -  \frac{\pi}{2} \,\big(\pa_a\c\big)\big(\pa^a\c\big)
\label{eps2d}
\ee
for two of the terms in (\ref{Sfluid}). Then integrating the relation (\ref{etachi}), so that 
\vspace{-2mm}
\be
\h = - \pi \c + \sdfrac{\th}{2} 
\label{etachirel}
\vspace{-2mm}
\ee
where $\th$ is a spacetime constant, and using (\ref{anom2}) for the chiral anomaly ${\cA}_2$ in $D\!=\!2$, we have 
\be
S_{\c\rm sf}\Big\vert_{D=2,\, \tA^a=0} = \int\!\! d^2x\! \left\{ - \frac{\pi}{2} \,\big(\pa_a \c\big)\big(\pa^a \c\big) 
- \c\, ^*\!F + \frac{\th}{2\pi}\, ^*\!F \right\} = S_{\rm eff}[\c; A] + \frac{\th}{2 \pi} \int\!\! d^2x\!\ ^*\!F 
\label{fl2d}
\ee

Thus the hydrodynamic superfluid action (\ref{Sfluid}), postulated on the basis only of {\it macroscopic} conservation laws for an irrotational and isentropic fluid
with the axial anomaly {\it is} the {\it microscopic} QFT effective action of the zero temperature Schwinger model (\ref{Seffchi}) 
up to a surface term, at zero coupling $e\!=\!0$. Note that the equivalence of the QFT to the chiral superfluid requires the addition of the (non-anomalous) $-\ve (\tn)$ 
energy density term in (\ref{eps2d}) to the effective action of the anomaly alone, for which the $\h$ variation of the first two terms of (\ref{Sfluid}) 
is sufficient to yield (\ref{anomJ5}). 

Since $^*\!F$ is a total divergence, addition of the constant $\th$ term does not affect 
the local dynamics. Instead this term is related to the topology of the gauge field configuration space. This topology, periodicity in $\th$ 
and hence $\h$ or $\c$ are made clear by careful treatment of the zero modes of the system in a finite linear spatial
volume $L$ in the real time Hamiltonian formulation of Appendix~\ref{Sec:Winding}. 

Since
\vspace{-6mm}
\be
[\h (t,x), \Pi_\h (t,x')]= \pi\, [\ch(t,x), \dot\ch(t,x')] = i\, \d(x-x')
\ee
the canonical commutation relation (\ref{etaPi}) of the superfluid hydrodynamic description coincides with the commutator
(\ref{comchi}) for the chiral boson of QED$_2$, which then implies the Schwinger term (\ref{j0j1com}) in the current commutators.
Since from (\ref{ep2}) $\tm/\tn = \pi$ is a constant in $D\!=\!2$, the wave eq.~of the hydrodynamic description
(\ref{etawave}) also coincides with that of the propagating chiral boson of $e=0$ electrodynamics (\ref{chieom}) in two dimensions.
Because of the simple relation (\ref{etachi}), the CDW solutions for $\c$ are also waves of the Clebsch potential $\h$,
which are the sound waves of the fluid. At finite equilibrium chiral density $\tn \neq 0$, these are CDW excitations of the Fermi 
surface at Fermi energy $\tm$. The vacuum limit $\tm \rightarrow 0, \tn \rightarrow 0$, with $\tm/\tn = \pi$ fixed shows that
this description holds when the Fermi surface of filled positive energy single fermion states goes over to the Dirac fermion
Fock vacuum where only the negative energy states of the Dirac sea are filled. In that limit the CDWs of the $\c$
field become chiral waves on the Dirac vacuum sea itself.

\section{Goldstone Theorem for Anomalous Symmetry Breaking}
\label{Sec:NambGold}

Since the axial anomaly with $\cA_D\neq 0$ breaks chiral symmetry explicitly rather than spontaneously, it may seem at first sight that Goldstone's 
theorem should not apply to ASB. However in the non-anomalous terms of the effective action (\ref{Sfluid}) and Hamiltonian (\ref{Hamcsf})
the velocity potential $\h$ appears only under derivatives, which therefore are left invariant by the constant shift symmetry $\h \rightarrow \h + \h_0$. 
Moreover in the anomaly term $\h$ multiplies $\cA_D= \pa_\l K^\l$, a total derivative, with $K^\l$ the Chern-Simons current. Hence the total 
action (\ref{Sfluid}) is also invariant under the constant phase shift transformation, up a topological term that does not affect the local dynamics.
Thus we should expect a variant of Goldstone theorem in QFT to guarantee this gapless mode of the chiral superfluid described by (\ref{Sfluid}) in a 
non-dynamical $A_\l$ gauge field background. In this section we show how Goldstone's theorem can be extended to the new case of symmetry 
breaking by the axial anomaly (ASB), by use of the canonical commutator (\ref{etaPi}) of the superfluid Hamiltonian, which is equivalent to the 
Schwinger terms in the anomalous commutator of currents.

The main element needed for a proof of Goldstone's theorem in the familiar case of SSB is the minimization of an effective potential at which some scalar
order parameter assumes a non-zero expectation value $\lag \F \rag \neq 0$ that is not invariant under the symmetry. In ASB, without assuming any
effective potential to be minimized, the $\h$ phase field does act as the natural order parameter of chiral symmetry breaking for the fermion theory.
We shall now show that provided
\be
\big\lag e^{2i\h}\big\rag = e^{2i\h_0}\equiv z_0 \neq 0
\label{expect}
\ee
is both well-defined and non-vanishing in the ground state of the system, the necessary and sufficient condition 
for the existence of a massless Nambu-Goldstone boson \cite{Nambu:1961tp,Nambu:1961fr,Goldstone:1962es} is satisfied. 

That $\lag e^{2i\h}\rag$ plays the role of a complex phase order parameter characterizing chiral symmetry breaking is made clear 
by integrating (\ref{etaPi}) over the spatial volume. This gives
\be
\big[\tQ(t), \h(t, \vx)\big] = \int\! d^d\vx' \left[\tJ^0(t, \vx'), \h(t,\vx)\right] = -i
\label{Q5eta}
\ee
which is a c-number, whose further commutators with $\tQ$ vanish.  Hence
\vspace{-2mm}
\bea
e^{i \a \tQ(t)}\h(t, \vx)e^{-i \a \tQ(t)}&=&\eta(t,\vx) + i\a\left[ \tQ(t), \h (t,\vx)\right] 
+ \sdfrac{(i\a)^2}{2} \Big[\tQ(t), \left[ \tQ(t), \h (t,\vx)\right]\Big]+ \dots\nn
&=& \h(t, \vx) + \a
\label{etashift}
\vspace{-4mm}
\eea
shifts $\h(t,\vx)$ by a constant $\a$ under a global chiral rotation of magnitude $\a$. Therefore
\be
e^{i \a \tQ(t)}e^{2i\h(t, \vx)}e^{-i \a \tQ(t)} = \sum_{n=0}^\infty \sdfrac{i^n}{n!}\, \big(2\h(t,\vx) + 2\a\big)^n = e^{2i\a}\,e^{2i \h(t, \vx)}
\label{Q5rot}
\ee
as expected for a $2\pi$-periodic phase field associated with the breaking of global chiral symmetry.

\vspace{-2mm}
\subsection{Lorentz Invariant Case}
\label{Sec:NGLorentz}
\vspace{-2mm}

If in addition to (\ref{expect}) the vacuum is both Lorentz and translationally invariant, it follows that
\vspace{-1mm}
\be
\int d^D\!x \ e^{ik\cdot (x-y)}\, \big\lag \cT \tJ^\l (x) \, e^{2i \h(y)}\big\rag = i k^{\l}F(k^2)
\label{GoldLor}
\ee
for some Lorentz invariant function $F(k^2)$. Then we have
\be
\frac{\pa}{\pa x^\l} \lag \cT \tJ^{\l}(x)\, e^{2i \h (y)} \rag = \d (x^0- y^0) \Big\lag \big[\tJ^0(x), e^{2i\h (y)}\big]\Big\rag = 2\,\d^D(x-y) \,z_0
\label{divJ5}
\ee
from the commutation relation (\ref{etaPi}), in the absence of gauge fields or limit of very weak coupling $e\!\to 0$ where the anomaly $\cA_D=0$.
Multiplying (\ref{GoldLor}) by $-ik_\l \rightarrow -\pa/\pa x^\l$, integrating by parts and making use of (\ref{divJ5}) yields then
\vspace{-2mm}
\be
2z_0 = k^2 F(k^2)\quad\Longrightarrow\quad F(k^2) = \frac{2z_0}{\,k^2}
\label{GoldstoneLorentz}
\vspace{-2mm}
\ee
and therefore
\vspace{-6mm}
\be
\int d^D\!y \ e^{ik\cdot (x-y)}\, \big\lag \cT \tJ^\l (x) \, e^{2i \h(y)}\big\rag =  2i\,\frac{k^\l}{k^2} \,\lag e^{2i \h}\rag
\label{GoldNam}
\ee
demonstrating the existence of the massless Nambu-Goldstone pole at $k^2 =0$ in any $D\!=\!2n$ even dimension 
where (\ref{etaPi}) holds due to the axial anomaly Ward Identity, provided $z_0 \neq 0$ in a Lorentz and translationally invariant state.

Thus the gapless acoustic mode of (\ref{etawave}), derived from the Hamiltonian and commutation relations of the canonical pair 
$\{\h, \Pi_{\h}\}$, is indeed a consequence of Goldstone's theorem, extended to this new case of ASB, symmetry breaking through the axial 
anomaly and anomalous Schwinger commutators. This establishes a microscopic QFT basis for superfluidity and a collective Goldstone sound 
mode in anomalous fermion systems.

We note that assuming the $\h$ phase has a well-defined expectation value in (\ref{expect}) is a statement about long-range order 
in the ground state of the system, where $\h$ is a composite or collective boson degree of freedom composed of fermion/anti-fermion pairs, as can be made 
explicit in $D\!=\!2$ by (\ref{etachirel}) and the fermion bosonization relations reviewed in Appendix \ref{Sec:Massless}. Rather than
the unmodified (non-anomalous) symmetry generators of SSB, we have employed the {\it anomalous} commutators required by the axial 
anomaly itself. The fact that (\ref{expect}) holds only approximately for a range of distance scales in $D\!=\!2$ where quasi-long-range order 
holds in the limit $e\!\to\!0$ but volume $L\!\to\!\infty$ is discussed in detail in Appendix \ref{Sec:Winding}, where the relation to the
fermion condensate and quasi-long-range-order is further explored. A gapless Goldstone mode related to the axial anomaly was discussed, 
in the context of universality of transport properties in chiral media in~\cite{Alekseev:1998ds}.\footnote{After this paper had been submitted 
for publication, we became aware of somewhat different considerations of Goldstone bosons in anomalous theories \cite{Delacretaz:2019}.}

\vspace{-2mm}
\subsection{Goldstone Sound Mode in General Case of Anomalous Superfluid Hydrodynamics}
\label{Sec:GenCase}
\vspace{-2mm}

The Goldstone mode propagates at the speed of light $c\!=\!1$ if and only if the ground state is Lorentz invariant as assumed in (\ref{expect})-(\ref{divJ5}). 
This need not be the case. Indeed if the wave eq.~(\ref{etawave}) for the phase field $\h$ in the superfluid description is linearized around its 
equilibrium solution
\vspace{-2mm}
\bes
\bea
\h&=&\tm\, t+\d\h \label{vareta}\\
\d \left(\frac{\tn}{\tm}\right) &=&  \frac{d}{\,d \tm} \left(\frac{\tn}{\tm}\right)\d \tm =  \frac{d}{\,d \tm} \left(\frac{\tn}{\tm}\right)\, \d\dot\h
\eea
\label{etapert}\ees
where $\d \tm = \d\dot\h\,$ follows from variation of (\ref{mu5eta}) and use of (\ref{vareta}) in the fluid rest frame. 
Then from the variation of (\ref{varJ}) with $A_{5\l} =0$ we obtain
\bes
\bea
\d \tJ^0 &=&  \tm\, \d\!\left(\frac{\tn}{\tm}\right) + \frac{\tn}{\tm}\, \d\dot \h = \left[\tm \frac{d}{d \tm\!}\! \left(\frac{\tn}{\tm}\right) + \frac{\tn}{\tm}\right]\d\dot \h
= \frac{d\tn}{d\tm}\ \d \dot \h \label{delJ50} \\
\d \tJ^i &=&- \frac{\tn}{\tm}\ \pa_i (\d \h)
\eea
\label{delJ}\ees
where the equilibrium (spacetime independent) values for $\tn, \tm$ and their derivatives are to be used. Eqs.~(\ref{delJ}) can be combined and
written in the covariant form
\be
\d \tJ^\l = \frac{d\tn}{d\tm}\left[u^\l u^\n \big(1- v_s^2\big) - v_s^2 \,g^{\l\n}\right] \pa_\n (\d\h)
\label{delJ5}
\ee
by use of the kinetic velocity $u^\l$ which is $\d^\l_{\ 0}$ in the fluid rest frame, and
\be
v_s^2 \equiv \frac{\tn}{\tm}  \frac{d\tm}{\,d\tn} = \frac{dp}{d\ve}
\label{vels}
\ee
by (\ref{mu5def}) and (\ref{pres}). Thus the wave eq.~(\ref{etawave}) linearized about equilibrium is
\be
\pa_\l \big(\d \tJ^\l\big) = \frac{d\tn}{d\tm}\left(\frac{\pa^2}{\pa t^2}- v_s^2 \,\na^2\right) \d\h = 0
\label{deletawave}
\ee
in the absence of the anomaly source, and the gapless Nambu-Goldstone mode propagates at speed $v_s \le 1$, becoming the speed of light
if and only if $\tm$ and $\tn$ are linearly related, and $p=\ve$.

Upon quantization the relations (\ref{Q5eta})-(\ref{Q5rot}) continue to hold for the variations $\d J^\l$ and $\d \h$. Thus (\ref{GoldLor}) may
be reconsidered for the linearized variations away from the ground state of the general non-vanishing $\tm, \tn$, and
\be
\int\! dt e^{-i\w (t-t')}\!\int  d^d\!\vx \ e^{i\bk\cdot (\vx-\vx')}\, \big\lag \cT \d \tJ^\l (t,\vx) \, e^{2i\d\h(t'\vx')}\big\rag = 
-i \left[u^\l u^\n \big(1- v_s^2\big) - v_s^2 \,g^{\l\n}\right] k_\n F(\w, |\bk|)
\ee
in terms of a scalar function $F(\w, |\bk|)$, which follows from (\ref{delJ5}) and the fact that the variation of the velocity potential $\d\h$ 
and $\tm, \tn$ are spacetime (pseudo)scalars. Repeating the steps leading to (\ref{GoldstoneLorentz}) leads then to 
\be
\left[-(k\cdot u)^2 \big(1- v_s^2\big) + v_s^2 \,k^2 \right] F(\w, |\bk|) = \big(- \w^2 + v_s^2\, \bk^2 \big) F\big(\w, |\bk|\big) = 2z_0
\ee
which implies
\vspace{-6mm}
\be
F\big(\w, |\bk|\big ) = \frac{2z_0}{- \w^2 + v_s^2\, \bk^2 }
\label{Goldvs}
\ee
instead, showing the gapless acoustic propagator pole in superfluid hydrodynamics for general $v_s$. 

An important point to notice about this derivation is that Lorentz invariance is spontaneously broken in general by a background $\tm$, 
and this is reflected in both the time dependence of (\ref{vareta}) and the necessity of taking the variation of the magnitude $\tn/\tm$ 
of the chiral current into account in (\ref{delJ50}), this variation being responsible for the sound speed $v_s$ differing in general from 
the speed of light, although the acoustic CDW Goldstone mode remains gapless. It is also instructive to consider the variation of 
the energy-momentum tensor (\ref{EMT}) linearized around a constant background $n_5$,
\be
M^{\l\n}\d J_{5,\n}\equiv \Bigg\{\left[g^{\l\n}+u^\l u^\n\left(1-v_s^2\right)\right]\left(u\cdot\pa\right) + u^\l\pa^\n-v_s^2\,u^\n \pa^\l\Bigg\}\d J_{5,\n}=0
\ee
which possesses non-trivial solutions in general $D\!=\! d+1$ dimensions only if the Fourier components of $\d J_{5,\a}$ satisfy the relation
\vspace{-3mm}
\be
\det M= (-i\w)^{d-1}\left[\w^2-v_s^2|\bk|^2\right]=0\,.
\vspace{-2mm}
\ee
This is the same gapless condition as (\ref{Goldvs}) for the Goldstone pole, showing that the only propagating mode in the non-dissipative 
anomalous superfluid is the gapless (first) sound mode.

\vspace{-2mm}
\section{The Axial Anomaly and Effective Action in Four Dimensions}
\label{Sec:ChiAnom4}
\vspace{-2mm}
\subsection{The Massless Pole in Four Dimensions}
\vspace{-2mm}

\begin{figure}[t]
\centering
\vspace{-1.2cm}
\includegraphics[width=.4\textwidth]{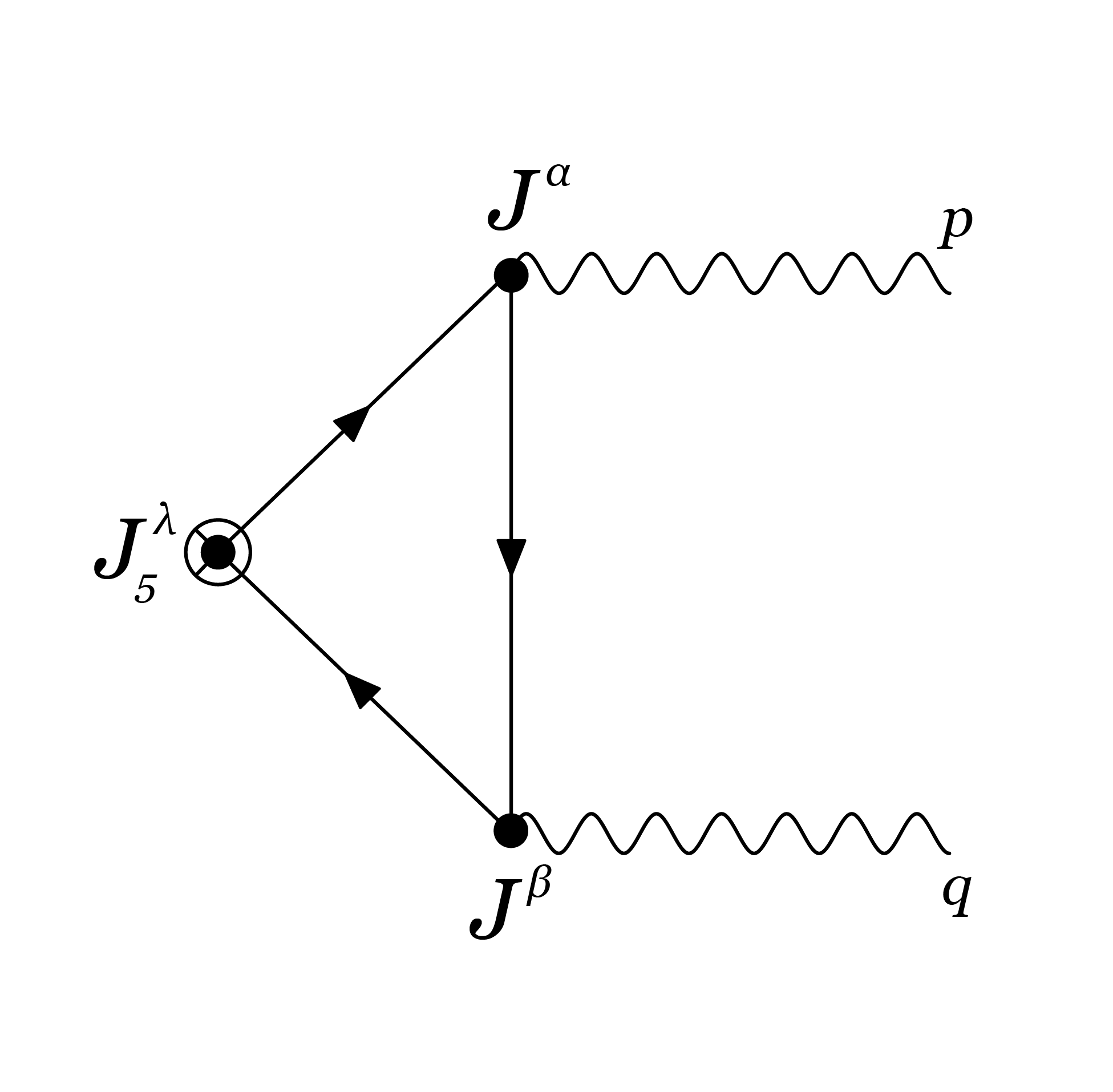}
\vspace{-1cm}
\caption{The Axial Anomaly Triangle Diagram Amplitude $\G^{\l\a\beta}(p,q)$}
\label{Fig:tri}
\vspace{-3mm}
\end{figure}

In $D\!=\!4$ dimensions the axial current $\tJ^\l = J_5^\l = \bar\psi \g^\l\g_5\psi$ has the anomalous divergence\hspace{-1mm}
\footnote{In $D\!=\!4$ we adhere to the more standard convention of multiplying the gauge potential $A_\l$ by the coupling $e$, hence 
the anomaly $\cA_4$ in (\ref{anomJ5}) by $e^2$, and relabel $\tg= \g_5, \tJ^\l = J_5^\l, \tm = \m_5,\tn= n_5$ {\it etc.}}
\bea
\pa_\l J^\l_5 = e^2\cA_4 = \frac{\a}{2\pi} \,F^{\l\n}\, ^*\!F_{\l\n} = \frac{2\a}{\pi}\,\vec {E\cdot B}
\label{axanom4}
\eea
for massless fermions, where $^*\!F_{\l\n} \equiv \frac{1}{2} \e_{\l\n\r\s}F^{\r\s}$ is the dual of the field strength tensor 
$F_{\l\n}$ and $\a = e^2/4 \pi$ \cite{Adler:1969gk,Adler:1969er,Bell:1969ts,treiman2015lectures}. The axial anomaly (\ref{axanom4}) 
in $D\!=\!4$ results from the one-loop triangle diagram of Fig.~\ref{Fig:tri}. In momentum space labeling $k= p +q$ 
the ingoing momentum at the axial vertex, and $p$ and $q$ the outgoing momenta on the photon legs, the triangle 
diagram may be evaluated explicitly
\be
\G^{\l\a\b}(p,q)=\sum^6_{i=1}f_i(k^2;p^2,q^2) \,\t_i^{\l\a\b}(p,q) 
\label{TriGam}
\ee
and expressed as a sum over six basis tensors $\t_i^{\l\a\b}(p,q)$ multiplied by scalar form factor functions $f_i$ of the three Lorentz invariants $k^2, p^2, q^2$.  
The coefficient functions $f_i$ are given {\it e.g.} in Refs.~\cite{PhysRev.129.2786,Giannotti:2008cv} and for zero fermion mass are 
\bes
\bea
&&f_1(k^2; p^2,q^2)=  f_4(k^2; q^2,p^2) = \frac{4\a}{\pi}\int^1_0 dx\int^{1-x}_0 dy\frac{xy}{\cD}   \\
&&f_2(k^2; p^2,q^2)= f_5(k^2; q^2,p^2)= \frac{4\a}{\pi}\int^1_0 dx\int^{1-x}_0 dy\frac{x(1-x)}{\cD} \\
&&f_3(k^2; p^2,q^2) =  f_6(k^2; q^2,p^2) = 0\label{f36}
\eea
\label{f12}\ees
where the denominator $\cD=(p^2x+q^2y)(1-x-y)+xy k^2$, in the basis where
\bes
\bea
&&\t_1^{\l\a\b}(p,q) =  \t_4^{\l\b\a}(q,p) = -p\cdot q\,\e^{\l\a\b\g} p_\g  -p^\b\,\y^{\l\a}(p,q)\\
&&\t_2^{\l\a\b}(p,q) =  \t_5^{\l\b\a}(q,p) = p^2\,\e^{\l\a\b\g} q_\g  +p^\a\,\y^{\l\b}(p,q)
\eea
\ees
and
\vspace{-9mm}
\be
\y^{\a\b}(p,q) \equiv \e^{\a\b\r\s}\,p_\r q_\s\,.
\ee
The two tensors $\t_3, \t_6$ are linearly dependent on the other four and redundant, and in any case unnecessary because of (\ref{f36}). 
The Feynman parameter integrals for $f_i$ in (\ref{f12}) can be evaluated in terms of digamma functions \cite{Armillis_2009}, but these 
explicit expressions will not be needed in the following.

If the current $J^\l_5$ is decomposed into its longitudinal and transverse components
\be
J^\l_5 = J^\l_{5\, \parallel} + J^\l_{5 \perp}\,,\qquad \qquad \pa_\l J^\l_{5 \perp} = 0
\ee
it is clear that only the longitudinal component $J^\l_{5\, \parallel}$ contributes to the anomalous divergence (\ref{axanom4}).
Similarly the triangle amplitude (\ref{TriGam}) may be decomposed into its longitudinal and transverse parts\be
\G^{\l\a\b}(p,q) = \G^{\l\a\b}_\parallel(p,q) +\G_\perp^{\l\a\b}(p,q)\,,\qquad\qquad k_{\l}\, \G_\perp^{\l\a\b}(p,q) =0 
\label{TriGam1}
\ee
so that the transverse part does not contribute to the anomaly. The longitudinal part explicitly exhibits a $1/k^2$ pole, 
and gives the total anomaly
\be
\G^{\l\a\b}_\parallel(p,q) = \frac{2\a}{\pi}\,\frac{k^\l}{k^2}\,\y^{\a\b}(p,q) \,,\qquad\qquad 
k_\l\G^{\l\a\b}(p,q) = k_\l\G^{\l\a\b}_{\parallel}(p,q) = \frac{2\a}{\pi}\,\y^{\a\b}(p,q)
\label{longGam}
\ee
which is (\ref{axanom4}) in momentum space. As in $D\!=\!2$ the appearance of the $1/k^2$ pole in (\ref{TriGam1}) 
signals that the anomaly is associated with a massless pseudoscalar collective excitation, here and in higher dimensions residing 
in the longitudinal subsector of the full theory. Whereas the form of the axial anomaly and hence the longitudinal sector
of the axial current with its $1/k^2$ pole is protected at higher loop orders by the Adler-Bardeen theorem \cite{Adler:1969er}, 
the transverse sector is not so protected.

Since the longitudinal projection of $J^\l_5$ is 
\vspace{-2mm}
\be
J^\l_{5\,\parallel}= \pa^\l \left(\sq^{-1} \pa_\n J^\n_5 \right)
\label{Jlong}
\vspace{-2mm}
\ee
the axial anomaly (\ref{axanom4}) corresponds to the non-local one-loop 1PI quantum effective action \cite{Giannotti:2008cv,Blaschke:2014ioa}\be
S_{\rm anom}^{NL}[A, A_5] \!=\! \frac{\a}{2\pi}\! \int\! d^4x\! \int \! d^4y\, [A^5_{\m}\,\pa^\m]_x \sq^{-1}_{xy}\,[F^{\l\n}\, ^*\!F_{\l\n}]_y
\label{SNL4}
\ee
where $\sq^{-1}_{xy}= \frac{1\,}{4 \pi^2}\, (x-y)^{-2}$ denotes the massless scalar propagator in $D\!=\!4$. The appearance of the massless 
$1/k^2$ pole (\ref{TriGam1}) and massless scalar propagator $\sq^{-1}_{xy}$ in (\ref{SNL4}), in the longitudinal sector of the $D\!=\!4$ axial 
anomaly is thus a simple kinematic consequence of the anomalous Ward Identity for the axial current. As in $D\!=\!2$ it leads to the expectation
that it is related to anomalous chiral symmetry breaking and Goldstone's theorem of Sec.~\ref{Sec:NambGold}, although the $\h$ phase field 
has not yet been identified, and will appear only in the local form of anomaly effective action of (\ref{Sanom}) below.
\vspace{-2mm}

\subsection{Local Effective Action and Anomalous Current Commutators in $D\!=\!4$}
\label{Sec:AnomCom}
\vspace{-1mm}

As in the $D\!=\!2$ case, the massless boson degree of freedom represented by the $1/k^2$ pole in (\ref{TriGam1}) or $\sq^{-1}_{xy}$ in (\ref{SNL4})
is a collective mode of a fermion pair intermediate state in the anomaly amplitude \cite{Giannotti:2008cv}. That it is also a CDW
may be made explicit by expressing the non-local action (\ref{SNL4}) in a local bosonic form. Since the non-local action (\ref{SNL4}) involves 
the axial potential $A^5_\l$ and $[F\, ^*\!F]$ asymmetrically, expressing this action in a local form apparently requires the introduction 
of two pseudoscalar fields ($\h, \c$), as suggested in \cite{Giannotti:2008cv}. Unlike in $D\!=\!2$, (\ref{Sanom}) is only part of the 1PI effective action 
of QED$_4$, with the dependence upon the transverse component $J_{5\perp}^{\l}$ not fixed by the anomaly. If two fields ($\h, \c$) are varied 
independently, a second massless wave eq.~would result, apparently implying the existence of {\it two} independent gapless modes. However 
this is not warranted by the triangle amplitude (\ref{TriGam}) itself where only a single $1/k^2$ pole appears. 

The previous $D\!=\!2$ example and chiral fluid action provides the way around this problem. Thus rather than introducing two scalar fields,
consider instead the local anomaly action
\be
S_{\rm anom}[\h; A,A_5]= \!\!\int\!\! d^4x\! \left\{\big(\pa_\l\h + A^5_\l\big)\,J_5^\l + \h \cA_4\right\}
\label{Sanom}
\ee
together with the variational principle that this effective action should be stationary against variations of the axial current $J_5^\l$. Then
$J_5^\l$ acts as a Lagrange multiplier field enforcing the constraint, $\pa_\l\h + A^5_\l= 0$, in the absence of any $\ve$ term added to 
(\ref{Sanom}). Solving this constraint for $\h$
\vspace{-2mm}
\be
\h= -\sq^{-1} \pa^\l A_\l^5 = -\sq^{-1} \pa^\l A_{\l\,\parallel}^5
\ee
\vspace{-2mm}
and substituting this back into (\ref{Sanom}) reproduces exactly the required non-local action (\ref{SNL4}) with its massless pole. Variation of 
$S_{\rm anom}$ with respect to $\h$ also reproduces the axial anomaly (\ref{axanom4}). The local action (\ref{Sanom}) does not require the 
introduction of a second scalar field, and this variational principle of $\d S_{\rm anom}/\d J_5^\l =0$ does not lead to a second independent 
massless mode, but (\ref{Sanom}) together with the variational principle for the axial current is completely equivalent to the effective action
(\ref{SNL4}) derived directly from the fermionic QFT axial anomaly.

That a single bosonic degree of freedom is described by (\ref{Sanom}) is made clear by defining the momentum canonically conjugate to $\h$
\vspace{-2mm}
\be
\Pi_\h = \frac{\d}{\d \dot\h } S_{\rm anom} = J_5^0
\vspace{-2mm}
\ee
as in (\ref{J05}), which then implies the equal time commutator
\be
\big[\h(t, \vx), \Pi_\h (t, \vx ')\big] = \big[\h(t, \vx), J^0_5(t, \vx ')\big] = i\, \d^3(\vx - \vx ')
\label{etaPi4}
\ee
upon quantization. Thus $\h$ and $J^0_5$ form a {\it single} canonical pair, and hence describe just a single 
gapless bosonic degree of freedom associated with the $U^{ch}(1)$ chiral anomaly in $D\!=\!4$, as in $D\!=\!2$.

Furthermore the electromagnetic current due to the axial anomaly may be found from
\be
J^\l = \frac{\d S_{\rm anom}}{\d A_\l\ } =\frac{\a}{2\pi} \frac{\d}{\d A_\l} \int\!\! d^4x\,  \h\, F^{\m\n}\, ^*\!F_{\m\n} = \frac{2\a}{\pi} \, 
\, ^*\!F^{\l\n} \, \pa_{\n}\h
\label{Jeta}
\ee
in terms of $\h$, which has the components
\vspace{-2mm}
\bes
\bea
&&J^0 =  - \frac{2\a}{\pi} \,{\vec B \cdot \na}\eta \label{J0eta}\\
&&{\vec J} = \frac{2\a}{\pi} \big({\vec B}\, \dot\eta - {\vec E \times\na} \eta  \big)\,.
\label{Jetadot}
\vspace{-2mm}
\eea
\label{Jveceta}\ees
Then making use of (\ref{etaPi4}), we have 
\bes
\bea
\big[J^0(t, \vx), J^0_5(t, \vx ')\big] &=& -\frac{2i\a}{\pi} \, {\vec B \cdot \na_{\vec x}} \, \d^3(\vx - \vx ') \label{J0J05}\\
\big[{\vec J}(t, \vx), J^0_5(t, \vx ')\big] &=& -\frac{2i\a}{\pi} \, {\vec E \times \na_{\vec x}}\, \d^3(\vx - \vx ')\hspace{5mm}
\eea
\label{JJcom}\ees
in a background electric or magnetic field \cite{Adler:1970qb,Gross:1970ee,treiman2015lectures}. Like the axial anomaly (\ref{axanom4}) itself, 
these current commutator Schwinger terms are anomalous, in the sense that they are apparently zero if the unregularized Dirac fermion anti-commutation 
relations are used. As in $D\!\!=2$ these Schwinger commutator terms in fermionic currents are in fact a consequence of the axial anomaly, and 
follow necessarily from the canonical commutator (\ref{etaPi}) of the bosonic effective action of the axial anomaly (\ref{Sanom}), which
therefore passes an important consistency check, showing that there is a single {\it bona fide} pseudoscalar collective degree of freedom 
which is not apparent at the classical level or the free Dirac theory, that is necessarily associated with the $U^{ch}(1)$ chiral anomaly also in $D\!=\!4$.

Let us emphasize that the Schwinger terms (\ref{JJcom}) in the current commutators depend only upon the longitudinal anomalous 
part of the triangle diagram, represented by (\ref{Sanom}).  Other commutators and in particular $\big[J^0(t, \vx), {\vec J}_5(t, \vx ')\big]$
which depend upon the transverse part of the amplitude $\G_{\perp}^{\l\a\b}(p,q)$ are not determined by $S_{\rm anom}$, not protected by the 
Adler-Bardeen theorem \cite{Adler:1969er}, and can be canceled by regularization scheme dependent `seagull' terms, hence removed 
entirely \cite{Adler:1970qb,treiman2015lectures}. The essential and unavoidable anomalous current commutators are (\ref{JJcom}), 
and these are entirely accounted for the local anomalous effective action (\ref{Sanom}), together with the canonical commutation relation 
(\ref{etaPi4}) it implies for a single {\it bona fide} bosonic degree of freedom.

\subsection{Chiral Magnetic and Separation Effects from the Anomaly Effective Action}

The axial anomaly effective action (\ref{Sanom}) succinctly incorporates several macroscopic chiral effects. Making use of (\ref{Jetadot}) in the case of a constant uniform $\vec B$ field, we find
\vspace{-2mm}
\be 
{\vec J} = \frac{2\a}{\pi} \,\m_5 \,\vec B
\label{CME}
\ee
in the Lorentz frame where $\vec\na \h = 0$, and $\dot \h = \m_5$ (changing notation $\tm= \m_5$ in $D\!=\!4$). This is the Chiral Magnetic Effect (CME), 
which has been discussed in the literature in various contexts \cite{Vilenkin:1980fu, Kharzeev:2007jp, Fukushima:2008xe, Buividovich:2009wi, Kharzeev:2009pj, Son:2009tf, Pu:2010as, Sadofyev:2010pr, Sadofyev:2010is, Kalaydzhyan:2011vx, Hoyos:2011us, Nair:2011mk, Son:2012wh, Stephanov:2012ki, Jensen:2012jy, Fukushima:2013, Khaidukov:2013sja, Avdoshkin:2014gpa}.

Since the longitudinal projection of the chiral current can be expressed as the pure gradient (\ref{Jlong}), one can define
$\c \equiv \sq^{-1} \pa_\n J^\n_5$, and express the axial anomaly in the form 
\be
\pa_\l J^\l_{5} = \pa_\l J^\l_{5 \parallel} = \sq \c = \cA_4 = \frac{2\a}{\pi} \,{\vec E \cdot \vec B}
\label{chieq4}
\ee
of a massless wave eq.~for the local $\c$ field describing the gapless bosonic mode, with the chiral anomaly as its source, just as in the $D\!=\!2$ 
case (\ref{chieom}). Analogously to $D\!=\!2$ this gapless mode is a collective mode of the two-fermion intermediate state in the
anomaly amplitude, if fermion masses and interactions can be neglected, and is a CDW in $J^\l_5$.
The components of the axial current expressed in terms of $\c$ are
\vspace{-8mm}
\bes
\bea
&&J^0_{5} = J^0_{5\parallel} = - \dot\c \label{J05chi}\\
&&{\vec J}_{5} = {\vec J}_{5\parallel} = {\bf \na} \c\,.
\label{J5ichi}
\vspace{-2mm}
\eea
\label{J5chi}\ees
In a static, constant ${\vec B} = B \bf \hat x$ field and parallel static electric field ${\vec E} = - \na \F = - {\bf \hat x} \,\frac{d\F}{dx}$ 
in the same direction, (\ref{chieq4}) becomes
\vspace{-6mm}
\be
\frac{d}{dx} \left(\frac{d \c}{dx} \right) = -\frac{2\a}{\pi} \frac{d \F}{dx}\, B
\ee
assuming also $\c = \c(x)$. Integrating this once and substituting into (\ref{J5ichi}) gives
\be
{\vec J}_5 = \na \ch  =  \frac{2\a}{\pi} \,\m\, \vec B
\label{CSE}
\ee
upon taking $\F= -\m$ for the charge chemical potential. In this way the Chiral Separation Effect (CSE) is also implied by and follows simply and
directly from the axial anomaly.  \footnote{Another example of the effect of anomalies is the Chiral Vortical Effect, see {\it e.g.}\cite{Son:2009tf,Sadofyev:2010is}, 
which is present in a rotating system of massless fermions even in the limit of zero charge when the anomalous divergence is zero.}

\vspace{-2mm}
\subsection{Anomalous Hall Effect}
\label{Sec:AHE}
\vspace{-2mm}

Yet another macroscopic effect which is succinctly captured by the anomalous effective action (\ref{Sanom}) is the Anomalous Hall Effect (AHE) in a Weyl semi-metal. 
A good model for a Weyl semi-metal is given by the Dirac theory with an external constant axial field $\vec A_5= \D \vec k$, with only spatial components 
corresponding to a shift $\D \vec k$ between the Weyl nodes in momentum space \cite{ZyuzinBurkov:2012,HosurQi:2013,goswami2013axionic}. In other words 
the fermionic spectrum is linear in momentum but the energies of right- and left-handed fermions reach zero at different points separated by $\vec A_5$.  In the presence of a constant $\vec A_5$ and a background electric field the current (\ref{Jetadot}) obtained by varying the anomaly effective action is
\vspace{-2mm}
\be 
{\vec J} =  -\frac{2\a}{\pi}{\vec E \times\na} \eta = -\frac{2\a}{\pi} \,\vec{A}_5\times\vec E
\vspace{-2mm}
\ee
where we use that $\vec{\na}\h=-\vec{A}_5$ corresponds to a constant axial field (\ref{Sanom}). This implies a linear phase shift in the
chiral field $\h = -\D\vec k\cdot \vec x$ in the effective action responsible for the AHE  \cite{ZyuzinBurkov:2012,HosurQi:2013,goswami2013axionic,Kim:2014}.

Thus the macroscopic CME, CSE, and AHE are all consequences of the same effective action (\ref{Sanom}), which is derived directly from and equivalent to the
bosonic form of the microscopic fermion QFT of the axial triangle anomaly (\ref{SNL4}) with its massless pole, on the one hand, and identical to the corresponding terms
in the general chiral superfluid effective action (\ref{Sfluid}) on the other. 

\section{Dimensional Reduction and Chiral Magnetic Waves}
\label{Sec:DimRed}

Unlike the case in $D\!=\!2$, the vacuum triangle amplitude (\ref{TriGam}) in massless QED$_4$ has both longitudinal and transverse parts. 
Since the anomaly action (\ref{Sanom}) takes account only of the longitudinal projection of the anomalous triangle diagram of massless fermions 
in QED$_4$, and provides no information about the transverse part of the chiral current, it is clearly incomplete. The action
(\ref{Sanom}) is also incomplete in that it contains no $J^\l_5$ dependence other than the minimal linear $(\pa_\l\h + A^5_\l)\,J_5^\l$ 
term, which results in the simple constraint $\pa_\l\h =- A_\l^5$, and hence no relation between the $\h$ potential and the chiral current, 
analogous to (\ref{etachi}) in $D\!=\!2$. It is that relation that enabled us to identify the propagating massless chiral boson $\c$ of the 
bosonized Schwinger model, satisfying (\ref{chieom}) with the Goldstone boson of the chiral superfluid description in (\ref{etawave}). 
This relation and the identity of the fluid action with that of the Schwinger model resulted from adding the non-anomalous energy density 
$-\ve(\tn)$ of (\ref{ep2}) to the effective action of the anomaly. 

In this section, we show that in the special case of a constant, uniform magnetic field background the transverse part of the anomaly amplitude 
(\ref{TriGam}) {\it vanishes}, and the four dimensional axial anomaly reverts to the $D\!=\!2$ case, and moreover with a simple completion 
of $-\ve(\tn)$, the CDW of dimensional reduction coincides with a Chiral Magnetic Wave (CMW) along the magnetic field direction.

Let $\vec B\!=\!F_{23}(0)\, \bf \hat x$ have only a zero momentum component in the $\bf \hat x$ direction with $A_{\b =2,3}(q)$ the 
corresponding gauge potential in the transverse $\bf \hat y$, $\!\bf \hat z$ directions as $q\!\rightarrow\! 0$. Computing $\G^{\l\a\b}(p,q)A_{\b}(q)$ 
in this limit, only the tensors $\t_1$ and $\t_2$ in (\ref{TriGam}) which are linear in $q$ contribute (as they are necessary to form a gauge invariant 
magnetic field source), but we can neglect $q$ and $q^2$ otherwise, setting $k = p + q \rightarrow p$ and $k^2=p^2$ in the denominator $\cD$ 
of (\ref{f12}). Thus $\cD=k^2 x(1-x)$ in this limit, and the Feynman parameter integrals (\ref{f12}) are trivially evaluated to give
\be
f_2=2f_1=\frac{2\a}{\pi k^2} \qquad {\rm for} \qquad q^2= 0\,.
\ee
Then taking ${\vec k}_\perp\!=\! {\vec p}_\perp \!= \!0$  in the transverse $\bf \hat y$, $\bf \hat z$ directions and noting that $k_\n F^{\n\l}(q)=0$, we find
\bea
\hspace{-1.5cm} \lim_{q\rightarrow 0}\G^{\l\a\b}(k-q,q)A_\b(q)\Big\vert_{\vec k_\perp=0} &=& \frac{2i\a}{\pi k^2}\left(k^2\d^\a_{\ \n}-k^\a k_\n\right)\tilde{F}^{\l\n}\nn
&=&\left\{ \begin{array} {cc}2i\a B\,\Pi_{2}^{ac}(k) \e^{\ b}_{c}\qquad & {\rm if} \qquad \a =a,\  \l = b \\
0 &{\rm otherwise} \end{array}\right.
\label{GAlim}
\eea
where $\a = a\,, \l = b$ range only over the $0, 1$ subspace of the $D\!=\!4$ spacetime and $\Pi_2^{ab}(k)$ is the $D\!=\!2$ vacuum polarization of (\ref{Pi2}). 
Thus, the full triangle diagram contracted with a constant uniform magnetic field reduces to the 2D anomalous self-energy polarization of Fig. \ref{fig:JJ}, 
and the $1/k^2$ pole in the 4D triangle anomaly becomes at ${\bf k}_\perp =0$ precisely the $1/k^2$ propagator pole of the effective boson 
$\c$ in the 2D two-point polarization tensor $\lag j^a j^c_5\rag$ of (\ref{Pitil2}). 

As a consistency check we may calculate the longitudinal part of the triangle amplitude (\ref{longGam}) directly. Contracting $\G_\parallel^{\l\a\b}(p,q)$ with 
$A^\b(q)$ and taking the same kinematic limit as in (\ref{GAlim}) we find
\vspace{-3mm}
\be
\lim_{q\rightarrow 0}\G^{\l\a\b}_\parallel(k-q,q)A_\b(q)\Big\vert_{{\vec k}_\perp = {\vec p}_\perp = 0}
=\lim_{q\rightarrow 0}\frac{2\a}{\pi}\frac{k^\l}{k^2}v^{\a\b}(p,q)A_\b(q)=-2i\,\frac{\a}{\pi}\,\frac{k^\l}{k^2}\,\tilde{F}^{\a\r}k_\r
\ee
where the only surviving indices are two-dimensional ranging over $t,x$.  Then $\tilde{F}^{ab}=\e^{ab} B$ for the 2D subspace of 
4D spacetime and the Schouten relation for $\e^{ab}$ ({\it cf.} Appendix \ref{Sec:Schouten}) results in
\be
\lim_{q\rightarrow 0}\G^{ca\b}_\parallel(p,q)A_\b(q)\Big\vert_{{\vec k}_\perp = {\vec p}_\perp = 0}
=\ \frac{2i\a B}{\pi k^2}\left(k^2\d^a_{\ b}-k^a k_b\right)\e^{cb}
\ee
which coincides with (\ref{GAlim}). This proves that the transverse part of the anomalous triangle diagram does not contribute 
in the dimensional reduction limit of a constant, uniform magnetic field, which is accounted for completely by its longitudinal
part and $1/k^2$ pole, which in this limit of $\bk_\perp=0$ becomes precisely the $1/k^2$ pole of the $D\!=\!2$ Schwinger
model of Sec.~\ref{Sec:Schw}. 

A third, and independent {\it non-perturbative} check of dimensional reduction is to make use of the polarization operator of
fermions in a constant, uniform magnetic field in the Lowest Landau Level (LLL) approximation
\vspace{-3mm}
\be
\big\lag \cT J^a(t,x,\by) J^b_5(t,x',\by')\big\rag_{_{\!\!LLL}}\hspace{-2mm} = 
2\a B \!\int\!\frac{d\w}{2\pi} \!\int\!\frac{dk}{2\pi}\!\int\!\frac{d^2{\bf k}_\perp}{(2\pi)^2}
\,e^{ik(x-x') + i{\bf k}_\perp \cdot (\by - \by')}\exp\!\left (\!-\frac{k_\perp^2}{2eB}\right)\Pi_2^{ac}(\w, k)\e_c^{\ b}  
\label{LLL}
\ee
given in terms of the $D\!=\!2$ polarization (\ref{Pi2}) with $a,b = t,x$ \cite{Loskutov1976, Calucci_1994, Gusynin:1994xp, 
Gusynin:1998zq, Fukushima:2011nu,Miransky:2015ava}. The exponential dependence of this expression upon $1/eB$
is obtained by treating the magnetic field background exactly, rather than in first order perturbation theory of (\ref{TriGam}).
Nevertheless, when we integrate (\ref{LLL}) over the transverse $\by$, thereby setting ${\bf k}_\perp\! = \!0$,
and evaluate the current commutator expectation
\be
\Big\lag \big[J^0(t,x,\by), J^0_5(t,x',\by')\big]\Big\rag_{_{\!\!LLL}}\hspace{-2mm} 
= \!\int\!\frac{d\w}{2\pi} \!\int\!\frac{dk}{2\pi}\!\int\!\frac{d^2{\bf k}_\perp}{(2\pi)^2}
\,e^{ik(x-x') + i{\bf k}_\perp \cdot (\by - \by')} \big\{2\, {\rm Im}\, \Pi_2^{01}(\w + i \e, k)\big\}
\label{JJPol}
\ee
from the imaginary part of (\ref{LLL}), we find
\bea
&&\int d^2 \by \Big\lag \big[J^0(t,x,\by), J^0_5(t,x',\by')\big]\Big\rag_{_{\!\!LLL}}\hspace{-2mm}
= 2\a B  \!\int\!\frac{d\w}{2\pi} \!\int\!\frac{dk}{2\pi}\,e^{ik(x-x')}\left\{2\, {\rm Im}\, \Pi_2^{01}(\w + i \e, k)\right\}\nn
&&\hspace{1.3cm}= 2\a B\, \Big\lag \big[j^0(t,x), j^1(t,x')\big] \Big\rag = -\frac{2i\a B}{\pi} \, \frac{\pa}{\pa x}\, \d(x-x')
\label{comLLL}
\eea
consistent with the anomalous commutators (\ref{J0J05}) derived from the $D\!=\!4$ triangle amplitude, simply proportional to the
$D\!=\!2$ current commutator (\ref{j0j1com}).

The fact that the LLL approximation saturates the anomalous commutator is consistent with the fact that only the LLL in a constant 
magnetic field has gapless excitations, so that if an external electric field is turned on adiabatically only fermions in the LLL can be excited, 
and the $D\!=4\!$ axial anomaly factorizes into its $D\!=\!2$ counterpart with a transverse density proportional to the magnetic field 
strength $B$ \cite{Nielsen:1983rb,Basar:2012gm}. 

Lastly, it is clear that in a constant uniform magnetic field with $\vec E (t,x)$ along the $\vec B$ direction, the four dimensional axial
anomaly (\ref{axanom4}) becomes a simple factor $2 \a B$ times the two dimensional axial anomaly (\ref{anom2}). In this
case the four dimensional ${\mathbb R}^4$ base manifold factorizes into ${\mathbb R}^2 \!\otimes\! {\mathbb R}^2$, and the
topology of the gauge field mapping the periodic $x$ domain to the gauge field configuration space ${\mathbb S}^1 \rightarrow {\mathbb S}^1$
applies just as in Appendix~\ref{Sec:Winding}, as does the Atiyah-Singer index theorem in $D\!=\!2$ \cite{Rupertsberger:1980fe}. Thus
we should expect all aspects of the previous $D\!=\!2$ analysis of superfluid CDW and Schwinger boson to carry over
directly to $D\!=\!4$ in a constant uniform magnetic field background with simple replacement of the $D\!=\!2$ dimensional
coupling $e^2$ by $2\a B$ in $D\!=\!4$ That this CDW Schwinger boson following directly from the anomaly effective action
(\ref{Sanom}) is in fact the Chiral Magnetic Wave (CMW) discussed in the literature by several authors \cite{Kharzeev:2010gd} may be seen 
as follows.

If in addition to the constant uniform magnetic field $B$ the massless fermions in the LLL are placed in a state of small but non-zero chiral
chemical potential $\m_5^2 \ll eB$, the relation between $n_5$ and $\m_5$ is linear, {\it cf. e.g.} \cite{Fukushima:2008xe}
\vspace{-6mm}
\be
n_5=\frac{e B\,}{2\pi^{2}}\,\m_5\,,\qquad \frac{n_5}{\m_5} = \frac{eB\,}{2\pi^2}
\label{LLLEoS}
\ee
corresponding to an energy density and pressure
\bes
\bea
&&\ve(n_5)=\int\! \m_5\,dn_5=\frac{\,\pi^2}{eB}\, n_5^2 \label{ener4B}\\
&&p(\m_5)=\m_5 n_5-\ve=\frac{e B}{4\pi^{2}}\,\m_5^2 = \ve \label{press4B}
\eea
\ees
respectively, This is in agreement with the form of the $D\!=\!4$ polarization operator in the LLL projection (\ref{LLL}),
which indicates that the system response to perturbations of gauge or axial fields is effectively two-dimensional 
if the fields are independent of the transverse spatial directions $\by$. Notice that the proportionality coefficient $2 \a B$
in (\ref{LLLEoS}) relative to the corresponding $D\!=2$ relation (\ref{ep2}) is exactly the same as that appearing in the
relative axial anomaly coefficients or anomalous current commutators (\ref{comLLL}) between two and four dimensions.
Thus if the energy density term (\ref{ener4B}) is appended to the anomaly effective action (\ref{Sanom}) to form the chiral
superfluid effective action (\ref{Sfluid}) in the same manner as in Sec.~\ref{Sec:equiv}, we obtain
\vspace{-2mm}
\be
J_5^\l+\frac{eB}{2\p^2}\, \pa^\l\h=0
\label{LLLconstr}
\vspace{-2mm}
\ee
by variation with respect to $J_5^\l$. Hence restricting to spatiotemporal variations in the $t,x$ components only, the CDW propagating along the 
magnetic field direction described by $\h$, generates not only a wave of electric density due to (\ref{Jetadot}) but also a wave of the 
axial density due to (\ref{LLLconstr}) and just coincides with the CMW of Refs.~\cite{Kharzeev:2010gd,Rybalka:2018uzh} found by other means.

This CDW/CMW also induces a small oscillating longitudinal electric field parallel to $\vec B$ by (\ref{J0eta}) and the Gauss law
\vspace{-4mm}
\be
{\na \cdot \bE} = - \frac{2\a}{\pi} \,{\bB \cdot \vec\na}\h \Longrightarrow \bE = - \frac{2\a}{\pi} \,\vec\na \left(\frac{1\ }{\vec{\na}^2}\!\right) {\bB \cdot \vec\na}\h = 
- \frac{2\a}{\pi} \h \vec B
\vspace{-2mm}
\ee
for slowly varying $\h (t,x)$, and assuming no transverse electromagnetic radiation. 
Thus the axial anomaly eq.~of motion for $\h$ (\ref{etawave}) with (\ref{LLLEoS}) becomes
\vspace{-2mm}
\be
\pa_\l J_5^\l = \frac{eB}{2\pi^{2}}\, \big(\pa_t^2-\pa_x^2\big)\,\h 
= \frac{2\a}{\pi}\,\vec {E\cdot B}= -\left(\frac{2\a}{\pi}\right)^{\!2} B^2\, \h
\label{etawave4B}
\vspace{-2mm}
\ee
and we find that just as in $D\!=\!2$ for the Schwinger boson, while the CDWs are massless in the absence of interactions 
with the electromagnetic field, they acquire a mass term $M^2 = 2 \a eB/\pi = e^{3}B/2\pi^{2}$ and satisfy the {\it massive} wave eq.
\vspace{-2mm}
\be
\left(\pa_t^2-\pa_x^2 + \frac{2\a}{\pi}\,eB\right)\h= 0
\vspace{-2mm}
\ee
when these interactions are taken into account, in agreement with the literature \cite{Kharzeev:2010gd,Rybalka:2018uzh}.
The speed of propagation of the CDW/CMW is also $v_s =1$ from (\ref{press4B}). Thus the macroscopic CMW
is here recognized to be a direct consequence of the axial anomaly and its anomalous massless pole, the Goldstone CDW sound mode
of anomalous chiral symmetry breaking, and a collective excitation of fermion/anti-fermion Cooper-like pairs, described by the effective action 
(\ref{Sanom}), becoming massive by its electromagnetic interactions, exactly as in the $D\!=\!2$ Schwinger model.

Since the role of the mass of the $D\!=\!2$ theory is taken by the substitution $e^2 \rightarrow 2 \a\, eB$, the full analysis of Appendix~\ref{Sec:Winding} 
applies, $2\h$ has $2 \pi$-periodicity, $z_0$ and the fermion chiral condensate are non-zero according to (\ref{z0expect}) and (\ref{condML}) in the limit 
$M\rightarrow 0, L\rightarrow \infty$, and the relativistic Goldstone Theorem of Sec.~\ref{Sec:NambGold} applies with $v_s=1$ in that limit. Since and 
$n_5/\m_5$ is again a constant in this case of constant, uniform $B$ field, we may again consider the limit $n_5, \m_5 \rightarrow 0$ with constant ratio. 
In that limit the CMWs are chiral waves on the Fermi-Dirac sea of the LLL ground state.

\section{Chiral Superfluid Hydrodynamics in Four Dimensions}
\label{Sec:ChiFlu4}

Since the anomaly action requires completion by some effective energy density $\ve$, we may also consider the case of
pure chiral density and no background magnetic field. For free fermions
\bes
\bea
&&n_5 = 2\! \int_{|\vec p| \le \m_5}  \frac{d^3 \vec p}{(2 \pi)^3} = \frac{\m_5^3}{3 \pi^2} = \left(\!-J_5^\l J_{5\,\l}\right)^{\!\frac{1}{2}}\\
&&\ve (n_5) =  2\! \int_{|\vec p| \le \m_5}  \frac{d^3 \vec p}{(2 \pi)^3}\,  |\vec p| = \frac{\m_5^4}{4 \pi^2} 
= \sdfrac{3}{4} \,(3 \pi^2)^{\frac{1}{3}}\, n_5^{\frac{4}{3}}=  \sdfrac{3}{4} \,(3 \pi^2)^{\frac{1}{3}}\, \left(\!-J_5^\l J_{5\,\l}\right)^{\!\frac{2}{3}}\\
&&\hspace{2cm}p(\m_5) = \m_5 n_5 - \ve =\frac{\m_5^4}{12 \pi^2} =\frac{\ve}{3}
\eea
\label{ener4}\ees
in $D\!=\!4$ spacetime dimensions. Adding this $-\ve(\tn)$ term,
\be
S_{\rm eff} = S_{\rm anom}[\h; A,A_5] - \int\! d^4 x\, \ve(n_5) = \int d^4x\! \left\{\big(\pa_\l\h + A^5_\l\big)\,J_5^\l + \h \cA_4 - \ve (n_5)\right\}
\label{Seff}
\ee
is the minimal effective action for the massless fermion system at finite $n_5$. 
We recognize that the effective action (\ref{Seff}) is exactly the chiral fluid action (\ref{Sfluid}) with
$\tm = \m_5, \tn= n_5$ in $D\!=\!4$, with the same consequences. In particular, the variation with respect to
$J_5^\l$ is now non-trivial and leads to 
\be
\pa_\l\h  +  A^5_\l - \left(\!\frac{d\ve}{\,d n_5}\!\right)\left(\! \frac{dn_5}{d J_5^\l}\!\right)=0 \qquad \Longrightarrow 
\qquad  J_5^\l= -\frac{n_5}{\m_5}\, \pa^\l\h = -\frac{\m_5^2}{3\pi^2}\, \pa^\l \h= 0
\label{J5eta4}
\ee
at $A^5_\l = 0$, with
\vspace{-7mm}
\be
\m_5^2 = - \pa_\l \h\, \pa^\l \h\,.
\ee
In effect, adding the $-\ve(\tn)$ in (\ref{Seff}) which depends on the {\it total} $n_5$, amounts to supplying a certain completion of the anomaly
action (\ref{Sanom}) in its transverse sector, which is justified when $\m_5 \neq 0$, and $\m_5$ is larger than any other dimensionful mass 
or energy scales in the system.

The eq.~of motion (\ref{etawave}) for $\h$ resulting from $S_{\rm eff}$ is
\be
\pa_\l J_5^\l  =\sdfrac{1}{3 \pi^2}\left[\frac{\pa}{\pa t} \left(\m_5^2\, \frac{\pa \h}{\pa t}\right) - \vec\na \!\cdot\! \Big(\m_5^2\, \vec\na\h\Big)\right] 
= \frac{\a}{2\pi} \,F^{\l\n}\, ^*\!F_{\l\n}
\label{etaeom4}
\ee
so that the boson field $\h$ which was introduced in \cite{Giannotti:2008cv} in order to express the non-local anomaly effective action (\ref{SNL4}) 
in local form, is identified here as the Clebsch potential of a dissipationless, irrotational chiral fluid in $D\!=\!4$ as well, when $\m_5 \neq 0$. 
Consistent with the condition that $\m_5$ be larger than any other scales, and the hydrodynamic approximation long wavelength 
dynamics near equilibrium generally, the solutions of (\ref{etawave4}) are strictly valid only for small amplitude perturbations 
$\vert\pa_\l (\d\h) \vert \ll \m_5$ of the equilibrium Fermi surface, {\it i.e.}~as shallow chiral waves on the Fermi sea. Thus the 
analysis of (\ref{etapert})-(\ref{deletawave}) of Sec.~\ref{Sec:GenCase} applies and (\ref{etaeom4}) becomes 
\be
\pa_\l (\d J_5^\l) = \frac{\m_5^2}{\pi^2}\left(\frac{\pa^2}{\pa t^2}- v_s^2 \,\na^2\right) \d\h = e^2 \cA_4 = \frac{\a}{2\pi} \,F^{\l\n}\, ^*\!F_{\l\n}
\label{etawave4}
\ee
restricted to its proper range of validity $|\vec\na \h|, |\dot \h| \ll \m_5$, and where
\be
v_s^2 = \sdfrac{dp}{d\ve} = \sdfrac{1}{3}
\label{vels4}
\ee
is the sound speed of the acoustic CDW in $D\!=\!4$. 

The sound speed $v_s <1$ reflects the fact that $ n_5/\m_5$ is not constant in $D\!=\!4$, as it is in $D\!=\!2$, and must be varied in (\ref{J5eta4}). 
This leads to a certain (non-anomalous) transverse contribution to the axial current perturbations, and the breaking of Lorentz
invariance, as in (\ref{delJ5}), while nevertheless preserving a gapless CDW solution, as (\ref{etawave4}) shows. It is not possible to extrapolate (\ref{etawave4}) 
directly to the vacuum state where $\m_5, n_5 \rightarrow 0$ without departing from the region of validity of linearized perturbations from the finite 
density background. For that reason the chiral superfluid description of massless fermions cannot be immediately extended to the Dirac vacuum in 
$D\!=\!4$, as is possible in the Schwinger model, where the constancy of $\tn/\tm$ in that $D\!=\!2$ case leads to the wave eq.~(\ref{etawave}),
equivalent to (\ref{chieom}), which is already linear. Whether a different completion of the effective action extending the bosonic description of 
fermionic pair excitations to $\m_5 = 0$ exists, describing CDWs on the Dirac sea in $D\!=\!4$ as well is an interesting open question. 

Since the CDW acoustic mode has the axial anomaly as its source in (\ref{etawave4}),  that implies
\be
\d\h= \frac{e^2}{2\m_5^2}\left(\frac{1}{\pa_t^2 - v_s^2\,\na^2}\right) \bE\cdot\bB
\ee
with the corresponding local variations 
\bes
\bea
&&\d J_5^0 = \frac{e^2}{2\pi^2}\left(\frac{1}{\pa_t^2 - v_s^2\,\na^2}\right) \frac{\pa}{\pa t} \left(\bE\cdot\bB \right)\\
&&\d \vec{J}_5 = -\frac{e^2}{2\pi^2}\left(\frac{v_s^2}{\pa_t^2 - v_s^2\,\na^2}\right) \vec\na \left(\bE\cdot\bB \right)
\eea
\label{delJ5_4D}\ees
of the axial charge density and current respectively. If these axial charge and current perturbations are substituted 
back into the action (\ref{Sfluid}) or (\ref{Seff}), expanded to second order around the background
\bea
S_{\rm eff}^{(2)}= \frac{1}{\pi^2} \int\! d^4 x\, \Big\{\m_5^2\,\big[\d\dot{\h}^2-v_s^2\, (\vec\na\d\h)^2\big]+ e^2 \d\h \,\bE\cdot\bB \Big\}
\eea
is the low energy gapless CDW effective action. The first two terms can be seen as the second order correction to the pressure $p$. 
This action can also be expressed in a non-local form
\bea
S_{\rm eff}^{(2)}=\frac{4\a^2}{\m_5^2} \int\!d^4x\!\int\! d^4y \ (\bE\cdot\bB)_x \left(\frac{1}{\pa_t^2 - v_s^2\,\na^2}\right)_{\!\!xy} (\bE\cdot\bB)_y 
\eea
coupling the anomaly source at two different spacetime points by the retarded interaction with the acoustic CDW propagator.
This interaction may have interesting consequences in chiral media.

The pseudoscalar collective mode satisfying (\ref{etawave4}) shows that bosonization of the Fermi surface extends to higher spacetime 
dimensions, as has been suggested earlier in the condensed matter literature \cite{haldane2005luttingers,PhysRevB.48.7790,PhysRevLett.72.1393}, 
although collective boson dynamics of the Fermi surface or Luttinger liquid behavior in spatial dimensions $d>1$ has not been related to the axial anomaly 
of massless fermions to our knowledge. The restriction to small amplitude, long wavelength perturbations, consistent with the hydrodynamic limit, is what 
permits treating the background Fermi surface as effectively flat, as it is in a single $d\!=\!1$ space dimension. In this limit superfluid behavior is again recovered.
This may allow some interesting applications to low temperature condensed matter systems with gapless fermions.

\vspace{-5mm}
\section{Summary}
\label{Sec:Sum}
\vspace{-2mm}

Since this paper has covered aspects of several different sub-fields relating the microscopic QFT axial anomaly to macroscopic effects and superfluidity,
it is worthwhile here for the convenience of the reader to gather and summarize the main results, with pointers to the Section and specific relations where those results
are established:
\begin{enumerate}[leftmargin=8mm, label={\bf (\arabic*)}]
\vspace{-3mm}
\item The action principle and Hamiltonian for an irrotational and dissipationless anomalous quantum chiral superfluid at zero temperature are given by (\ref{Sfluid})
and (\ref{Hamcsf});
\vspace{-3mm}
\item In $D\!=\!2$ spacetime dimensions this action and Hamiltonian is completely equivalent to the fermionic Schwinger model of QED$_2$
in the limit of vanishing electric charge $e\!\to\!0$, {\it cf.} (\ref{fl2d});
\vspace{-3mm}
\item The gapless sound mode of superfluid hydrodynamics is both a Charge and Chiral Density Wave (CDW), a fermion/anti-fermion (or particle/hole) pair excitation of the 
Fermi surface at zero temperature and finite fermion density, which coincides with the Schwinger boson;
\vspace{-3mm}
\item In $D\!=\!2$ the superfluid description can be extended to zero fermion density, so that the acoustic CDW becomes 
a chiral wave on the Dirac sea and the Dirac vacuum itself may be viewed as a kind of superfluid medium;
\vspace{-3mm}
\item Goldstone's theorem can be extended in a novel way to chiral symmetry breaking through the axial anomaly, Anomalous Symmetry Breaking (ASB), 
(\ref{GoldNam}) or (\ref{Goldvs}), by making use of the anomalous Schwinger commutator of the chiral phase $\h$, with its canonical momentum
(\ref{etaPi4});\hspace{-5mm}
\vspace{-3mm}
\item A new local bosonic form of the effective action of the fermionic triangle anomaly in $D\!=\!4$ is given by (\ref{Sanom}),  together with a variational principle consistent
with the hydrodynamic (superfluid) effective action (\ref{Sfluid});
\vspace{-3mm}
\item Macroscopic quantum effects such as the CME, CSE and AHE all follow directly from this anomalous effective action (\ref{Sanom});
\vspace{-3mm}
\item In a constant uniform $\bB$ field with parallel $\bE$ field independent of transverse directions, the axial anomaly factorizes and the massless boson CDW reduces 
to that of the $D\!=\!2$ Schwinger boson, with the CDW along the $\bB$ field direction a Chiral Magnetic Wave (CMW)\cite{Kharzeev:2010gd,Rybalka:2018uzh}, 
which is therefore a direct consequence of the massless anomaly pole (\ref{TriGam1}),  in an explicit realization of dimensional reductions;
\vspace{-3mm}
\item In $D\!=\!4$ with non-zero chiral density $n_5$, $\d\h$ satisfies a gapless wave equation (\ref{etawave4}) of a
long wavelength CDW excitation of the Fermi sea, with a sound velocity (\ref{vels4}) $v_s < c$, realizing previous conjectures
of bosonization of the Fermi surface in $d >1$ spatial dimensions \cite{haldane2005luttingers,PhysRevB.48.7790,PhysRevLett.72.1393};
\vspace{-2mm}
\item The prediction of a gapless boson collective excitation being generated by the axial anomaly itself may be testable in weakly self-interacting
Dirac and Weyl semi-metals.
\vspace{-3mm}
\end{enumerate}

We have given a detailed account of the consistency of the superfluid description and Schwinger model in the special case of $d=1$ spatial dimension
in Sec.~\ref{Sec:Schw} and Appendices \ref{Sec:Massless} and \ref{Sec:Winding}, where quasi-long-range order applies. The Goldstone
theorem of Sec.~\ref{Sec:NambGold} remains valid for an arbitrarily large range of distances and times $|x-x'| \ll 1/e$ (\ref{xkrange}) in the limit
$e\!\to\!0, L\!\to\infty$ with $eL> 1$ fixed, whereas long range order and the fermion condensate vanishes, in the $L\!\to\! \infty$ limit 
with any finite $e \neq 0$ fixed, consistent with the Mermin-Wagner-Coleman theorem. 

Since $\tn/\tm$ is not a constant in $D > 2$ dimensions, and there is a transverse component in the chiral current which is not
fixed by the axial anomaly, the chiral superfluid description of massless fermions applies only to a subsector of the theory,
and is clearly incomplete. Nevertheless in this sector the axial anomaly pole (\ref{TriGam1}) exists, and expanding around 
non-zero chiral density and chemical potential $\d\h$ still satisfies a gapless wave equation (\ref{etawave4}) of 
long wavelength CDW excitation of the Fermi sea. 

The fact that the CDW  shape fluctuations predicted by the axial anomaly and the effective action (\ref{Seff}) are gapless raises several other interesting 
questions about the breaking of chiral symmetry in massless QED$_4$ in relation to the Goldstone theorem. 
The infrared divergences encountered in perturbation theory of massless QED$_4$ \cite{Gribov,Rubakov84,Morchio,PhysRevD.33.1755}, 
suggest that massless QED$_4$ does break chiral symmetry, exhibit fermion confinement and develop a non-zero fermion condensate, whose phase 
would then be related to $\h$. The considerations of the present work suggest that the axial anomaly and ASB rather than SSB may be the
mechanism by which this occurs. It would be instructive to carry out the calculation of the anomalous triangle diagram of Fig.~\ref{Fig:tri} in massless QED$_4$ 
at finite chiral chemical potential and chiral density $n_5$ to check explicitly the appearance of the gapless mode and CDW propagator appearing in (\ref{delJ5_4D}), 
and thus derive (\ref{etawave4}) and the effective action (\ref{Seff}) at finite $\m_5$ from first principles. 

Finally we remark that the theoretical considerations of this work may find concrete realization in recent discoveries of condensed 
matter systems with gapless fermion spectra, such as Dirac and Weyl semi-metals \cite{RevModPhys.83.1057,Armitage2018,HosurQi:2013}. 
The chiral anomaly and the effective action (\ref{Sanom}) is critical to the Anomalous Hall Conductivity 
and response of these systems to external fields \cite{ZyuzinBurkov:2012,HosurQi:2013,goswami2013axionic,Kim:2014}. It has also been 
suggested that Weyl semi-metals can support dynamic axionic excitations \cite{li2010dynamical, WangZhang:2013, Gooth:2019lmg}, which for
weak fermion-fermion self coupling may be identified with the Nambu-Goldstone mode of chiral symmetry breaking due to the axial anomaly discussed 
in this paper. The mechanism of massless fermion pairing due to the anomaly and the corresponding anomalous pole provides a possible microscopic 
description of these emergent phenomena. This will require extension of the theory to include finite temperature corrections and additional fermion-fermion
self-interaction terms. The important features of the superfluid state described in this paper could be probed by interactions with photons 
which should indicate the presence of a gapless pseudoscalar mode with axionic coupling, manifested also by the effective 
non-local vertex of the light-by-light scattering in the material \cite{Saxena}. 

\vspace{4mm}
\centerline{\bf Acknowledgments}

The authors are grateful to P. Glorioso, D. Kharzeev, A. Saxena, I. Shovkovy, and A. Stergiou for useful comments. A. S. would like particularly to 
thank V. I. Zakharov for many useful discussions. The work of A. S. is partially supported through the LANL LDRD Program. A. S. is also grateful 
for support by RFBR Grant 18-02-40056 at the beginning of this project.
\vspace{-4mm}

\bibliographystyle{apsrev4-1}
\bibliography{alpsbib}

\vspace{-5mm}
\appendix

\section{Energy-Momentum Tensor Conservation and Comparison of $\cH$ and $T^{00}$}
\label{Sec:Schouten}
\vspace{-3mm}

We note first that the Hermitian Dirac chiral matrix $\tg$ is defined in $D$ dimensions by 
\be
\tg =  \frac{i^{\frac{D}{2} -1}}{D!}\, \e_{\a_1\dots \a_D}\, \g^{\a_1}\dots \g^{\a_D}\ \stackrel{D=2n}{=} 
\  \frac{i^{n -1}}{(2n)!}\, \e_{\m_1\n_1\dots \m_n\n_n}\, \g^{\m_1}\dots \g^{\n_n}
\label{gamtilD}
\ee
where the totally anti-symmetric Levi-Civita tensor $\e^{\m_1\n_1\dots \m_n\n_n}$ satisfies the Schouten identity 
\be
g^{\l\b} \e^{\m_1\n_1\dots \m_n\n_n} + g^{\l\m_1}  \e^{\n_1\m_2\dots \n_n \b} + \dots + g^{\l\n_n}  \e^{\b \m_1\dots \n_{n-1}\m_n}=0
\label{Shou}
\ee
in which the sum is over all $2n+1$ cyclic permutations of the indices $(\b, \m_1, \n_1\dots, \m_n,\n_n)$. Since the tensor on
the left side of (\ref{Shou}) is totally anti-symmetric in all $D+1$ indices, but no such tensor exists in $D\!=\!2n$ dimensions, it must
vanish identically. 

To demonstrate the equivalence of (\ref{nconsA}) and (\ref{nconsJ}), as expected in a background electromagnetic field,
we make use of general form of the axial anomaly (\ref{anomJ5}) in $D$ (even) dimensions. Multiplying (\ref{Shou}) 
by $c_n F_{\m_1\n_1} \dots F_{\m_n \n_n}$ where $c_n = 2/(4\pi)^n n!$ is the coefficient in (\ref{anomJ5}) gives
\be
g^{\l\b} \cA_D + c_n \, D \, g^{\l\m_1}  \e^{\n_1\m_2\dots \n_n \b} F_{\m_1\n_1} \dots F_{\m_n \n_n} = 0
\label{anomShou}
\ee
since the latter $D$ terms are all the same after relabeling indices. Contracting with $\pa_\b \h$ then yields
\bea
- \cA_D\, \pa^\l \h &=& c_n\, D\,  g^{\l\m_1} \e^{\n_1\m_2\dots \n_n \b} F_{\m_1\n_1} \dots F_{\m_n \n_n} \pa_{\b} \h\nn
&=& c_n\, D\,  \e^{\m_1\n_1\dots \m_n\n_n} F^\l_{\ \ \m_1} \dots F_{\m_n \n_n} (\pa_{\n_1} \h)
\label{calAeta}
\eea
after another relabeling of indices. On the other hand from (\ref{Janom}),
\bea
J^{\m_1} &=& c_n \, D\, \e^{\m_1\n_1\dots \m_n\n_n}\, \pa_{\n_1} \big(\h\, F_{\m_2\n_2}\dots F_{\m_n\n_n}\big)\nn
&=& c_n \, D\, \e^{\m_1\n_1\dots \m_n\n_n}\, F_{\m_2\n_2}\dots F_{\m_n\n_n}\, (\pa_{\n_1}\h)
\eea
by the Bianchi identity for the electromagnetic field strength, and therefore
\be
F^{\l\m}J_{\m} = F^\l_{\ \ \m_1}J^{\m_1}= c_n \, D\,\e^{\m_1\n_1\dots \m_n\n_n}\, F^\l_{\ \ \m_1} F_{\m_2\n_2}\dots F_{\m_n\n_n}\, (\pa_{\n_1}\h)
\label{FJ}
\ee
which coincides with (\ref{calAeta}). Thus the equivalence of (\ref{nconsA}) and (\ref{nconsJ}) is demonstrated.

The fact that the simple perfect fluid form of the Energy Momentum Tensor (\ref{EMT}) satisfies the partial conservation
law (\ref{nconsJ}) shows that it already contains the $J^\n A_\n$ interaction term of the underlying fermionic theory.
However, comparing its $T^{00}$ component with the canonical Hamiltonian $\cH$ of (\ref{Hamcsf}), one
finds that they differ by the anomaly term $\h \cA_D$. The reason for this difference is again the special form
of the axial anomaly (\ref{anom2}) being linear in the $F_{i0}$ electric component of the field strength tensor
in any dimension (all of the other indices of $F_{ij}$ being necessarily spatial by anti-symmetry of the $\e$ symbol). 

The difference between $T^{00}$ and $\cH$ is best illustrated by a simple model with one degree of freedom defined by the Lagrangian
\be
L_{tot}[\c, A]= \sdfrac{m_1}{2} \,\dot \c^2 + \c \dot A + \sdfrac{m_2}{2} \,\dot A^2 \equiv L_{\c} + L_A
\label{Lmodel}
\ee
where $L_\c$ comprises the first two $\c$ dependent terms, including the $\c \dot A$ interaction linear in the velocity of $A$.
This gives rise to a `current' $J= \d S_{tot}/\d A = -\dot \c$.
If $A(t)$, modeling the external gauge potential, is taken to be an arbitrary non-dynamical external field, its Lagrangian
$L_A$ and eq.~of motion can be neglected, and we may compute the Hamiltonian for the $\c$ field only, obtaining
\be
H_{\c} = \dot \c \frac{\pa}{\pa \dot \c} L_{\c} - L_\c = \frac{m_1}{2} \dot \c^2 -\c \dot A  =  \frac{1}{2 m_1}\, p_\c^2 -\c \dot A
\label{Hchi}
\ee
which contains the time dependent interaction, and yields the correct $\c$ eq.~of motion $m_1 \ddot \c = \dot A$,
analogous to the axial anomaly eq.~(\ref{chieom}) or (\ref{chieq4}) in the arbitrary potential $A(t)$.

On the other hand, being linear in the velocity $\dot A$, the interaction term will {\it not} appear in the covariant definition
of the energy since the action  $\int dt \, \c \dot A$ is invariant under arbitrary reparameterizations of time. If one
further calculates the total Hamiltonian corresponding to (\ref{Lmodel})
\bea
H_{tot} &=&  \dot \c \,\frac{\pa}{\pa \dot \c} \,L_{tot} +  \dot A \,\frac{\pa}{\pa \dot A}\, L_{tot} - L_{tot} = H_{\c} + \c \dot A + \sdfrac{m_2}{2} \dot A^2\nn
&=&  \sdfrac{m_1}{2} \dot \c^2  + \sdfrac{m_2}{2} \dot A^2
\eea
the interaction term apparently {\it cancels} entirely, when $H_{tot}$ is expressed in terms of the coordinate velocities, although of course
the conjugate momentum $p_A = m_2 \dot A + \c = const.$, and the eqs.~of motion consisting of 
\vspace{-8mm}
\be
m_1 \ddot \c = \dot A\,,\hspace{1cm} {\rm and} \hspace{1cm} m_2 \ddot A = - \dot \c
\label{Addot}
\ee 
still reflect the presence of the interaction. Since
\be
\frac{d}{dt}\left(  \sdfrac{m_1}{2}\, \dot \c^2\right) = \dot A \dot \c = - \frac{d}{dt}\left(\sdfrac{m_2}{2}\, \dot A^2\right)
\ee
the total Hamiltonian is conserved $\dot H_{tot} = 0 $ upon using both eqs.~of motion. The $\dot A \dot \c = EJ$ partial conservation
of the $m_1 \dot\c^2/2$ apparently `free' kinetic term of $\c$ is analogous to the $F^{\l\n}J_\n$ term in the partial conservation
(\ref{nconsJ}) of the covariant $T^{\l\n}$, for the apparently `free' matter, which is cancelled only if the Maxwell Eqs.~of the full interacting theory
are considered. Thus $m_1 \dot\c^2/2$, corresponding to $\int d^d x\,T^{00}$ of the covariant tensor (\ref{EMT}), contains the {\it full} interaction 
energy, without the $\c\dot A$ term, whereas the partial canonical Hamiltonian (\ref{Hchi}), corresponding to $\int d^d x\, \cH$ of (\ref{Hamcsf})
which differs from it and does contain the $\c \dot A$ term is not the true energy of even the $\c$ subsystem, although it does give the 
correct eq.~of motion (\ref{Addot}) in an external potential $A(t)$. 

This curious situation is clearly tied to the special nature of the interaction linear in velocity $\dot A$ which models the anomaly 
term $\cA_D$ in the fluid effective action (\ref{Sfluid}). If one takes $m_1= \tm/\tn = \pi$ and $m_2 = 1/e^2$ corresponding to
the $D\!=\!2$ Schwinger model, constrained to its spatially independent mode, one obtains from (\ref{Addot}) 
oscillatory solutions at the frequency $e/\sqrt{\pi}$, corresponding to the mass of the Schwinger model boson,
with the $p_A$ constant of motion proportional to the integration constant $E_0$ corresponding to constant background 
electric field in $E= e^2 \c + E_0$, corresponding in turn to the $\th$ parameter $\th = 2 \pi E_0/e^2$ of the model.

\vspace{-4mm}

\section{Massless Fermions in $D\!=\!2$, Bosonization and Chiral Charge Density States}
\label{Sec:Massless}
\vspace{-3mm}

We collect in this Appendix for completeness and the benefit of the reader some facts about massless fermions at zero temperature 
in $D\!=\!2$, and their bosonization which are used in the text. 

The two-component Dirac field $\psi$ for $1+1$ dimensional massless fermions can be written 
\be
\psi = \left( \begin{array}{c} \psi_+\\ \psi_- \end{array}\right) 
\ee
in terms of its chirality $\pm$ right or left moving components which can be treated separately in the absence of fermion mass. 
These components can be expanded in a Fourier series 
\be
\psi_{\pm} (x,t) =\frac{1}{\sqrt{L}} \sum_{q \in {\mathbb Z} + \frac{1}{2}}  c_q^{(\pm)}\, e^{ i k_q (x\mp t) }
\label{Fourier}
\ee
obeying anti-periodic boundary conditions on the spatial interval $x \in [0,L]$ with 
\be
k_q=\frac{2\pi q}{L}\,, \qquad q = \pm\sdfrac{1}{2}, \pm\sdfrac{3}{2}, \dots
\ee
$q$ taking on all (positive or negative) half-integer values, and the coefficients obeying the anti-commutation relations
\vspace{-8mm}
\be
\aco{c_{q'}^{(\pm)\dag}}{c_q^{(\pm)}} = \d_{qq'}
\label{anticom}
\ee
with all other anti-commutators vanishing. We employ a condensed notation (slightly different from that of Ref. \cite{Blaschke:2014ioa}), 
which is related to the more common notation by $c_q^{(+)} = b_q^{(+)}, c_q^{(-)\dag} = d_q^{(-)}$ denoting annihilation operators 
for positive energy modes if $q > 0$, and $c_q^{(+)} = d_{-q}^{(+)\dag}, c_q^{(-)\dag} = b_{-q}^{(-)\dag}$ denoting creation operators for positive 
energy modes if $q < 0$. Thus the free fermion Dirac vacuum is defined by
\vspace{-6mm}
\bes
\bea
&&c_q^{(+)}\big\vert 0\big\rag = c_q^{(-)\dag}\big\vert 0\big\rag=0\,,\qquad q= +\sdfrac{1}{2}, +\sdfrac{3}{2}, \dots\\
&&c_q^{(+)\dag}\big\vert 0\big\rag  = c_q^{(-)} \big\vert 0\big\rag = 0\,, \qquad q= -\sdfrac{1}{2}, -\sdfrac{3}{2}, \dots
\eea
\label{bdvac}\ees
\indent Fermion chiral densities can be expressed in the form
\be
:\!\psi^\dag_\pm \psi^{\ }_\pm\!: \ = \frac{1}{L} \sum_{n \in \Z} \r_n^{(\pm)}\, e^{-ik_nt}\, e^{\pm ik_n x}\,, \qquad k_n 
= \frac{2\pi n}{L}\,,\qquad \r^{(\pm)}_{-n} =  \rho^{(\pm)\,\dag}_n 
\label{jpsipm}
\ee
where the colons denote normal ordering with respect to the Dirac vacuum. Paying due attention to the distinction between creation 
and annihilation operators as in (\ref{bdvac}), we have 
\vspace{-1mm}
\be
\r^{(\pm)}_n \equiv  \sum_{q\in\Z + \inv2}:\! c^{(\pm)\dag}_{q}c_{q\pm n} ^{(\pm)}\!\!: = 
\left\{\begin{array}{l}\sum_{q>0} c_{q}^{(+)\dag}c_{q+n}^{(+)} -\sum_{q<0} c_{q+n}^{(+)}c_q^{(+)\dag} \qquad {\rm for} \quad \r^{(+)}_n\\
\sum_{q<0} c_{q}^{(-)\dag}c_{q-n}^{(-)} -\sum_{q>0} c_{q-n}^{(-)}c_q^{(-)\dag}   \qquad {\rm for} \quad \r^{(-)}_n\ \end{array}\right.
\label{rhodef}
\vspace{2mm}
\ee
with $n$ taking on integer values. Defining the normalized fermion bilinears
\be
a^{(\pm)}_n\equiv \frac{\ i}{\!\!\sqrt{|n|}}\ \r^{(\pm)}_n\,, \qquad n \neq 0
\label{andef}
\ee
a short calculation then shows that these obey {\it bosonic} canonical commutation relations
\be
\big[a^{(\pm)}_n, a^{(\pm) \dag}_{n'}\big]  =  \d_{n,n'}
\label{cancomm}
\ee
with the other commutators vanishing. As a result local bosonic field operators may be defined by
\vspace{-4mm}
\be
\phi_{\pm} (t,x)\equiv \frac{1}{2\pi} \sum_{n=1}^{\infty}\,\,\frac{\,1}{\!\!\sqrt{n}}\left(a^{(\pm)}_{n}e^{-ik_n(t\mp x)}
+a^{(\pm)\dagger}_{n}e^{ik_n(t\mp x)}\right) + \phi^0_\pm(t, x) \equiv \bar\phi_{\pm} + \phi^0_\pm(t, x)
\label{phiposn}
\ee
where $\phi^0_\pm$ is the contribution of the $n=0$ modes. These are linear in $t$ and $x$, and given by
\be
\phi^0_\pm= \inv{2\pi}R_\pm  + \frac{1}{L}(t\mp x)\,Q_{\pm}\,, \qquad Q_{\pm} \equiv \int_0^L dx \, :\!\psi^\dag_\pm \psi_\pm\!: =  \r_0^{(\pm)}
\label{zeromodes}
\ee
with their coefficients also operators obeying the commutation relation
\be
[R_{\pm}, Q_{\pm}] = i
\label{RQcom}
\ee
while the remaining commutators $[R_{\pm}, Q_{\mp}] = 0$. The chiral boson fields $(\c,\h)$ of the text are
\be
\c= -\big(\phi_+ -  \phi_-\big)\,,  \qquad  \h = -\pi\, \c = \pi\, \big(\phi_+ -  \phi_-\big)
\label{chietaphi}
\ee
in terms of the bosonic operators (\ref{phiposn})-(\ref{zeromodes}), explicitly expressing the bosonization in $D\!=\!2$.
The total electric and chiral charge operators defined by the Dirac vacuum normal ordering (\ref{jpsipm}) are 
\bes
\bea
Q &\equiv& \int_0^L dx\, j^0  = \int_0^L dx\, \frac{\pa\c}{\pa x} =\big( \phi^0_- - \phi^0_+\big)\Big\vert_{x=0}^{x=L} = Q_+ + Q_- \\
\tQ &\equiv& \int_0^L dx\, j_5^0  = -\int_0^L dx\, \frac{\pa\c}{\pa t} =\int_0^L dx\, \big( \dot\phi^0_+ -\dot \phi^0_-\big)= Q_+ - Q_-
\label{Q5def}
\eea
\ees
respectively, where (\ref{vecaxvec}), (\ref{phiposn}) and (\ref{chietaphi}) have been used. 

At zero temperature the lowest energy fermion state with finite chiral charge density is attained by filling single particle positive chirality 
states and single anti-particle negative chirality states up to the Fermi momentum $k\le k_F = \tm$. Thus we define the filled Fermi level states 
\be
\big\vert N\big\rag \equiv
\left\{\begin{array}{cl}\hspace{1mm}\prod\limits_{1/2 \le q \le q_{_N}}\hspace{2mm} c_{q}\,^{\!\!\!(+)\dag}\, c_{q}\,^{\!\!\!(-)} \big \vert 0 \big\rag 
&\ {\rm for} \quad N= 1, 2,\dots \\
\vspace{3mm}\qquad\quad\vert 0 \rag &\ {\rm for} \quad N=0  \\
\hspace{-4mm}\prod\limits_{\quad q_{_{N+1}} \le q \le -1/2}\!\!c_q\,^{\!\!\!(+)}\, c_q\,^{\!\!\!(-)\dag}\big \vert 0 \big\rag
&\ {\rm for} \quad N=  -1, -2, \dots  \end{array}\right.
\quad q_{_N}\equiv N - \sdfrac{1}{2} \le \frac{\tm L}{2\pi} < q_{_{N+1}}
\label{mu5state}
\ee
for every integer $N \in \mathbb{Z}$. For any real $\tm$, $N$ is the largest integer not greater than $\tm L/2 \pi + 1/2$.
The $\vert N\rag$ states are eigenstates of both electric and chiral charge:
\be
Q_{\pm} \big\vert N\big\rag = \pm N \big\vert N \big\rag\,, \quad\qquad Q\big\vert N\big\rag = 0\,, 
\quad\qquad \tQ\big\vert N\big\rag = 2N \big\vert N\big\rag
\label{QqM}
\ee
since there are equal integer numbers of occupied single particle and anti-particle states in $\vert N\rag$, and particles and anti-particles
carry opposite electric charges. 

Relations (\ref{zeromodes})-(\ref{RQcom}) also allow us to identify the $n\!=\!0$ mode operator of the $\h$ field
\be 
\hat\h_0 \equiv \inv{2}\, \big(R_+ - R_- \big)   \,,\qquad [\hat\h_0, \tQ] = i
\label{eta0}
\ee
whose conjugate momentum is the net chiral charge $\tQ$. Thus if $\hat\h_0$ is sharply defined, $\tQ$ is maximally uncertain, and conversely,
$\hat \h_0$ is completely uncertain in the eigenstates (\ref{QqM}) of fixed chiral charge. Since from (\ref{QqM}) the spectrum of $\tQ/2$ is 
isomorphic to the angular momentum operator $L_z$ on a circle, its conjugate $2 \hat\h_0$ is an angular phase variable with period $2 \pi$ 
on ${\mathbb S}^1$. Then defining 
\hspace{-3mm}
\be
U_0 \equiv e^{2 i \hat\h_0} = \exp \Big\{i \big(R_+-R_-\big)\Big\} 
\label{U0def}
\vspace{-2mm}
\ee
we observe that this unitary operator satisfies
\pagebreak
\be
\big[\tQ, U_0\big] = i \Big[\tQ, \big(R_+ -R_- \big)\Big] U_0 = 2  \, U_0\,,\qquad \big[Q, U_0\big]= 0
\label{Q5raise}
\ee
so that $U_0$ raises the chiral charge $\tQ$ by $2$ units, and its phase can be chosen so that
\be
U_0\vert N\rag = \vert N+1 \rag
\label{Uraise}
\ee
without changing the value of $Q\!=\!0$ in the vanishing electric charge sector. In the angular coordinate representation where 
$\hat \h_0\vert \h_0\rag = \h_0\vert \h_0\rag$, the eigenstates (\ref{QqM}) are represented
\be
\lag \h_0 \vert N\rag = \frac{1}{\!\sqrt{\pi}\,}\, e^{2iN \h_0}\,,\qquad \vert \h_0\rag = 
\frac{1}{\!\sqrt{\pi}\,} \sum_{N \in {\mathbb Z}} e^{-2 i N\h_0} \vert N\rag\,,\qquad
U_0\vert \h_0\rag = e^{2 i \h_0}\vert \h_0\rag
\label{Ndeta}
\ee
and $\tQ$ is represented as $-i\, \pa/\pa \h_0$. The normalization of the states is fixed by $\int_0^\pi  d\h_0\, |\lag \h_0 \vert N\rag|^2 = 1$. 

Mixed fermion bilinears are expressed in terms of the boson field $\h$ of (\ref{chietaphi})-(\ref{eta0}) and $U_0$ by 
\be
\psi^\dag\,_{\!\!\!\pm}\psi^{\ }_{\mp} = \frac{1}{L}\, :\!e^{\pm 2i \bar\h}\!: \, e^{\pm 2 i \hat\h_0} 
= \frac{1}{L}\, :\!e^{\pm 2i \pi (\bar\phi_+ - \bar\phi_-)}\!: \, U_0^{\pm 1}
\label{bilin}
\ee 
where the colons now denote normal ordering with respect to the zero mass bosons of $\bar\phi_{\pm}$ in (\ref{phiposn}) 
\cite{Klaiber:1967jz,Coleman:1974bu,WolfZittartz85,Hetrick:1988yg}. From (\ref{Uraise}) and (\ref{bilin}) it follows that the only 
non-zero matrix elements $\lag N'\vert\bar \psi \psi\vert N\rag$ are those with values of $N' = N \pm 1$, and in particular
\be
\big\lag N +1 \big\vert \psi^\dag\,_{\!\!\!+}\psi^{\ }_- \big\vert N \big\rag = \frac{1}{L} \, \exp\left\{2\pi i\, \big(2N+1\big) \sdfrac{t}{L} \right\} 
\label{condN}
\ee
which is independent of $x$. 

From the definition of $q_N$ in (\ref{mu5state}) we have
\be
\frac{\tm L}{2 \pi} + \frac{1}{2} = N + {\rm fr}\!\left\{ \frac{\tm L}{2 \pi} + \frac{1}{2}\right\}
\label{fracpart}
\ee
where fr\{...\} $\in [0,1)$ denotes the fractional part of the quantity within the brackets. 
Dividing $\tQ$ in (\ref{QqM}) by the linear volume $L$ and passing to the infinite $L$ limit
\be
\tn = \lim_{L\rightarrow \infty} \frac{1}{L}\,\big\lag N\big\vert \tQ\big\vert N\big\rag 
= \lim_{L\rightarrow \infty} \frac{2}{L} \left(\frac{\tm L}{2\pi} + \frac{1}{2} - {\rm fr}\!\left\{ \frac{\tm L}{2 \pi} + \frac{1}{2}\right\} \right) = \frac{\tm}{\pi}
\label{n52D}
\ee
is the chiral charge density in the continuum limit, since the fractional part drops out in this limit.

At exactly zero electric coupling $e=0$, $\tQ$ is conserved and any of its eigenstates $\vert N\rag$ are also
eigenstates of the free Dirac fermion Hamiltonian, 
\be
\hspace{-1mm}H_f \!=\! \int_0^L \!dx :\!\psi^{\dagger}\left(\!-i \s_3\frac{\pa}{\pa x}\right) \psi\!: \quad {\rm with} \quad 
H_f   \big\vert N\big\rag =\bigg(2 \! \sum_{q= \frac{1}{2}}^{q_{_N}} k_q\bigg) \big\vert N\big\rag =
\frac{2\pi}{L} N^2 \big\vert N\big\rag = \frac{\pi}{2L} \tQ^2\, \big\vert N\big\rag 
\label{Enerf}
\ee
so that the energy density of this state in the infinite volume limit is
\be
\ve =  \lim_{L\rightarrow \infty}  \frac{2}{L}  \sum_{q= \frac{1}{2}}^{q_{_N}} k_q 
= 2\int_{0}^{\tm} \frac{dk}{2\pi}\, k  = \frac{\tm^2}{2\pi} = \frac{\pi}{2}\, \tn^2\,.
\label{energy2D}
\ee
For comparison the fermion energy-momentum-stress tensor is defined by
\be
T_{\l\n} = -\sdfrac{i}{4}\, \Big[\bar\psi\,, \g_{(\l}\stackrel{\leftrightarrow\ \,} {\pa_{\n)}}\psi\Big]
\label{EMST}
\ee
where the commutator anti-symmetrizes the fermion operators. Taking the expectation
value of $T_{tt}$ gives $\ve$ in (\ref{energy2D}). while for $T_{xx}$ we obtain 
\be
p = 2 \int \frac{dk}{2 \pi} \,k  =   \frac{\tm^2}{2\pi} = \ve 
\label{press2D}
\ee
as required by the tracelessness of the stress tensor $-T_{tt} + T_{xx} = 0$ for a massless fermion. From (\ref{n52D})- (\ref{press2D}) we verify that
the average Grand Potential density for the massless fermion fluid $D\!=\!2$ is
\vspace{-5mm}
\be
\lim_{L\rightarrow \infty} \frac{\lag \W_f\rag}{L} =  \lim_{L\rightarrow \infty} \frac{\lag H_f -\tm \tQ \rag }{L} = \ve - \tm \tn = - p
\label{gibbs}
\ee
which is the Gibbs relation. Finally since the time-space component of the tensor (\ref{EMST}) generates 
Lorentz boosts, we may compute
\vspace{-5mm}
\be
\int_0^L\! dx:\!T^{tx}\! \!:\,\big|\th; \tm\big\rag = 0
\ee
in the finite $\tm$ state of (\ref{mu5vac}), thus proving that despite its time dependence the state $\vert\th; \tm\rag$ is Lorentz invariant, 
and the gapless Goldstone mode propagates at the speed of `light' $v_s= c\!=\!1$ for all $\tm$ in $D\!=\!2$ dimensions, consistent 
with the discussion in Sec.~\ref{Sec:NGLorentz}, for zero fermion mass.

\vspace{-4mm}
\section{Vacuum Periodicity and Fate of the Goldstone Mode in $D\!=\!2$\hspace{-3cm}}
\label{Sec:Winding}
\vspace{-3mm}

This appendix reviews some features of the Schwinger model, which are
collected here for completeness, particularly as they relate to the superfluid description of Secs.~\ref{Sec:Ideal}, \ref{Sec:ChiFlu2}, 
and the proof of the Goldstone theorem for ASB of Sec.~\ref{Sec:NambGold}. The extension to finite chiral density
in eqs. (\ref{mu5vac})-(\ref{condmu}) has not to our knowledge appeared previously. The fully solvable $D\!=\!2$ case is also quite illuminating in 
showing that ASB can share features with, but is nevertheless different from SSB.

In Sec.~\ref{Sec:ChiFlu2} we have shown that the superfluid effective action (\ref{Sfluid}) coincides with that of the Schwinger model,
QED in $D\!=\!2$ dimensions, in the limit of vanishing electric coupling $e\!\rightarrow\!0$. In that limit the Schwinger boson
has zero mass and is a gapless CDW and Goldstone boson of the superfluid description. On the other hand, it is known 
that Goldstone's theorem fails in $D\!=\!2$ \cite{Mermin1966,Coleman:1973}, at least in the limit $L\!\rightarrow\!\infty$ of infinite
spatial volume. In this section we also examine and resolve this apparent conflict by reconsidering the Hamiltonian form of the Schwinger 
model for finite $L$, with finite electric coupling $e$, then carefully taking the limits $e \!\rightarrow \! 0, L\!\rightarrow\!\infty$. 

The necessary preliminaries of quantization of free massless Dirac fermions in $D\!=\!2$ are reviewed in Appendix \ref{Sec:Massless}.  
At exactly $e\!\equiv\!0$ and any finite $L$ one can construct states labeled by $| N \rag$ in (\ref{mu5state}) which are filled up to the chiral Fermi level and 
are eigenstates of chiral charge $\tQ=2N$ with zero electric charge. Since their energy (\ref{Enerf}) is proportional to $\tQ^2$, hence $N^2$, 
the ground state of the free fermion system is the $N\!=\!0$ state with $\tQ \!=\!0$. Exact chiral symmetry is preserved. 
Then it follows from the fact that $\tQ$ and the phase operator $\hat\h_0$ are conjugate variables, {\it cf.} (\ref{eta0}), that the phase is 
completely uncertain in the eigenstates $|N\rag$ of chiral charge. Hence the condition $z_0 \neq 0$ for the Goldstone theorem in 
Sec.~\ref{Sec:NambGold} does not apply  in this case, since $z_0=\lag e^{2i\hat\h_0}\rag \sim \int_0^\pi d\h_0\, e^{2i\h_0}= 0$. 

As soon as the fermions are coupled to the gauge field, the bare fermion eigenstates of $\tQ$ states are no longer acceptable, because
they are not invariant under topologically non-trivial large gauge transformations defined by $\L(L)- \L (0) = 2 \pi n, n\neq 0$. As a 
consequence they also violate the cluster decomposition property, which signals the appearance of off-diagonal long range order 
and chiral symmetry breaking \cite{Lowenstein:1971fc,Kogut:1974kt}. This is another consequence of the anomaly.

The analysis of the limit $e \!\rightarrow\! 0$ is delicate, and is most conveniently carried out in the Hamiltonian Schr\"odinger picture in a finite
spatial interval $x \in [0,L]$ \cite{Manton:1985jm,Hetrick:1988yg,Link:1990bc}. The spatial component of the gauge potential $A_x$
can be decomposed into a longitudinal piece, $\pa_x \L$ and a `transverse' piece, denoted simply by $A$. In one spatial dimension, 
`transverse' means $\pa_x A = 0$ so that $A= A(t)$ is independent of $x$ and a single quantum mechanical variable. 
Defining a field $\Phi(t,x) \equiv \dot \L -A_t$ with the temporal component $A_t$, the general gauge potential can then be written 
\vspace{-4mm}
\bes
\bea
&&A_x =  A + \pa_x \L\\
&&A_t =  - \F + \dot\L
\eea
\ees
so that under the $U(1)$ gauge transformation
\vspace{-2mm}
\be
A_a \rightarrow A_a + \pa_a \l \Longrightarrow \L \rightarrow \L + \l
\vspace{-2mm}
\ee
and $\L$ parameterizes the gauge orbit. The remaining field $\F$ and single degree of freedom $A$ are gauge invariant,
as is the electric field $E= F^{01} = -\dot A - \pa_x \F$. Since the physical states must be fully gauge invariant, including
under topologically non-trivial gauge transformations, $\L = 2 \pi x n/L, n \in {\mathbb Z}$, under which $A \!\rightarrow\! A + 2 \pi n/L$,
it is clear that there is a periodic vacuum structure in which states with the Chern-Simons number
\be
N_{_{CS}}  \equiv \frac{1}{2\pi} \int_0^L\! dx\, A  = \frac{AL}{2\pi}
\label{NCS}
\ee
differing by an integer should be identified as gauge copies of each other.

If the $\th$ term of (\ref{fl2d}) is added to the Lagrangian of (\ref{Schwmod}) the momentum conjugate to $A$ is
\be
P_A \equiv  \frac{\d S_{cl}}{\d \dot A} = \frac{\dot A L}{e^2}- \frac{ \th L}{2 \pi}
\ee
while $P_\F=0$, so that $\F$ is non-dynamical and satisfies the Gauss Law constraint
\vspace{-2mm}
\be
\pa_x E_x = -\pa_x^2 \F = e^2 j^0 = e^2\, \psi^\dag \psi = e^2 \,\pa_x\c
\label{Gauss}
\vspace{-2mm}
\ee
where the last relation follows from (\ref{j0chi}). This leads to the Coulomb interaction term
\be
H_c  = \frac{1}{2} \int_0^L dx\, \F j^0  = \frac{e^2\!}{2} \int_0^L dx \int_0^L dx' \,j^0 (x)\, D_L(x,x')\, j^0 (x')
\label{Hc}
\ee
in the Hamiltonian, with $D_L$ the Green's function 
\be
D_L(x,x') = \frac{1}{L} \sum_{n \in {\mathbb Z},\, n\neq 0} \frac{1}{k_n^2}\, \exp\big\{ i k_n(x-x')\big\} 
= \frac{L}{12} - \frac{|x-x'|}{2} + \frac{(x-x')^2}{2L}\,, \quad x,x' \in [0,L]
\label{DLdef}
\ee
of the $d\!=\!1$ spatial Laplacian $-\pa_x^2$ defined on the periodic interval $x,x' \in [0,L]$, satisfying 
\be
-\pa_x^2\, D_L(x,x') =  \frac{1}{L} \sum_{n \in {\mathbb Z},\, n\neq 0}  \exp\big\{ i k_n(x-x')\big\}  = \d_L (x-x') - \frac{1}{L}
\label{Lapl}
\ee
with $k_n= 2 \pi n/L$. Integrating (\ref{Gauss}) over the spatial interval $[0,L]$ implies that the total electric charge must vanish 
on the physical state space with periodic boundary conditions. This insures that the $n\!=\!0$ constant term omitted from the 
sums in (\ref{DLdef}) and (\ref{Lapl}) does not contribute to $H_c$.  

The total Hamiltonian for the Schwinger model with general $e\!\neq\!0$ may then be written
\vspace{-2mm}
\be
H = \frac{e^2}{2L} \left(P_A + \sdfrac{ \th L}{2 \pi}\right)^2 + H_f(A) + H_c
\label{SchwHam}
\vspace{-2mm}
\ee
where $P_A = - i d/dA$ in the Schr\"odinger representation and
\vspace{-2mm}
\be
H_f(A) = -\int_0^L dx\, \psi^{\dag}(x)\, \s_3\, \big(i\pa_x + A\big) \psi (x)
\label{Hamf}
\vspace{-2mm}
\ee
is the Dirac fermion kinetic Hamiltonian depending on $A$. Substituting (\ref{Fourier}) into (\ref{Hamf}) gives
\vspace{-2mm}
\be
H_f (A) = \sum_{q \in {\mathbb Z} + \frac{1}{2}} \left\{ \big(k_q-A\big)\, c_q^{(+)\dag} c_q^{(+)}   -  \big(k_q -A\big)\, c_q^{(-)\dag} c_q^{(-)}  \right\}
\vspace{-2mm}
\ee
for the unregularized Dirac Hamiltonian. Regularization of $H_f(A)$ may be performed in a number of different ways, for example 
with the normal ordering prescription for the shifted Dirac vacuum
\bes
\bea 
&&c_q^{(+)} \big\vert 0\big\rag = c_q^{(-)\dag} \big\vert 0\big\rag = 0 \,, \qquad k_q-A \ge  0\\
&&c_q^{(-)} \big\vert 0\big\rag = c_q^{(+)\dag} \big\vert 0\big\rag = 0 \,, \qquad k_q-A <  0
\eea \label{cvac}
\ees
re-ordering and discarding the contribution to the energy of the Fermi-Dirac sea. This gives
\vspace{-2mm}
\bea
&&\hspace{-5mm} :\!H_f (A)\!:\  = \frac{2 \pi}{L} \left\{ \sum_{q  \ge N_{_{CS}} } \big(q- N_{_{CS}}\big)\, c_q^{(+)\dag} c_q^{(+)}   
+ \sum_{q < N_{_{CS}} } \big(N_{_{CS}}-q\big)\,  c_q^{(+)} c_q^{(+)\dag}\right.\nn
&&\left.+  \sum_{q \le  N_{_{CS}} } \big(N_{_{CS}} - q\big)\, c_q^{(-)\dag} c_q^{(-)}  \
+  \sum_{q  >  N_{_{CS}} } \big(q -N_{_{CS}}\big)\,  c_q^{(-)}c_q^{(-)\dag}  \right\}
\vspace{-2mm}
\eea
where the colons denote that normal ordering subtraction has been performed.

According to the definition of the fermion Fock vacuum state (\ref{cvac}), we generalize the definition of the $Q=0$ state with unequal chiral 
Fermi surfaces of (\ref{mu5state}) to arbitrary finite $A$ by \cite{Manton:1985jm} 
\vspace{-2mm}
\be
\big\vert N;A\big\rag \equiv
\left\{ \begin{array}{cl}\hspace{-4mm}\prod\limits_{\ N_{_{CS}}-1/2 < q \le q_{_N}} \!\!c_{q}\,^{\!\!\!(+)\dag}\, c_{q}\,^{\!\!\!(-)} \big \vert 0 \big\rag 
&\ {\rm for} \quad N=\big\lfloor N_{_{CS}} + 1/2\big\rfloor + 1,  \big\lfloor N_{_{CS}} + 1/2\big\rfloor + 2,\dots \\
\qquad\vert 0 \rag &\ {\rm for} \quad N=\big\lfloor N_{_{CS}} + 1/2\big\rfloor   \\
\hspace{-5mm}\prod\limits_{\quad q_{_{N+1}} \le q < N_{_{CS}}+1/2}\!\!\!\!\!\!\!c_q\,^{\!\!\!(-)\dag}\, c_q\,^{\!\!\!(+)}  \big \vert 0 \big\rag
&\ {\rm for} \quad N= \big\lfloor N_{_{CS}} + 1/2\big\rfloor -1, \big\lfloor N_{_{CS}} + 1/2\big\rfloor-2, \dots  \end{array}\right.
\label{mu5Astate}
\vspace{-2mm}
\ee
with $q_{_N}$ is in (\ref{mu5state}), and $\lfloor x \rfloor \in {\mathbb Z}$ denotes the largest integer not greater than $x$ 
($\lfloor x \rfloor \le x$ is called the `floor' of $x$). For $A=0, N_{_{CS}}=0$, this reduces to the free fermion case of (\ref{mu5state}).

Since the energies of the single particle occupied states are shifted: $k_q \rightarrow k_q - A$, the normal ordering also
shifts the $Q_{\pm}$ operators so that only those single particle states with positive energy
\vspace{-2mm}
\be
k_q - A > 0 \, \Longrightarrow q \ge \sdfrac{1}{2} + N_{_{CS}} 
\label{Ashift}
\vspace{-2mm}
\ee
should be counted in the sums for the renormalized charge operators $Q_\pm$ after the shift. Since
\vspace{-2mm}
\be
\sum_{q\ge \frac{1}{2} + N_{_{CS}}}^{q_{_N}} \!\!1 = \left(\sum_{n=1}^{N}  -  \sum_{n=1}^{N_{_{CS}}}\right)1
= N- N_{_{CS}}
\vspace{-2mm}
\ee
the eigenvalues of $Q_{\pm}$ in the state defined by (\ref{mu5state}) but with energy levels shifted by $A$ according to (\ref{Ashift}) are given by
\vspace{-7mm}
\bes
\bea
&&Q_{\pm}\big\vert N;A\big\rag  =\sum_{q\ge \frac{1}{2} + N_{_{CS}}}^{q_{_N}} \big\vert N;A\big\rag = \pm \big(N-N_{_{CS}}\big)
\big\vert N ;A\big\rag \\
&&\tQ\big\vert N;A\big\rag = (Q_+ - Q_-)\big\vert N;A\big\rag = 2\,\big(N -N_{_{CS}} \big)\big\vert N;A\big\rag
\label{Q5shift}
\eea
\ees
which shows that the chiral charge $\tQ$ receives a contribution from the gauge field $A$ relative to (\ref{QqM}) 
Likewise the state $\big\vert N;A\big\rag$ is also an eigenstate of the Dirac kinetic Hamiltonian with eigenvalue
\bea
&&E_{N;A} = 2\!\!\!\sum_{q\ge \frac{1}{2} + N_{_{CS}}}^{q_{_N}}\!\!\left( k_q - A\right) = \frac{4 \pi}{L} 
\left(\sum_{n=1}^{N}  -  \sum_{n=1}^{N_{_{CS}}}\right)\Big(n- \sdfrac{1}{2} -N_{_{CS}}\Big)\nn
&&= \frac{2\pi}{L} \Big(N(N+1) - N_{_{CS}}(N_{_{CS}}+1)-  (2N_{_{CS}}+1)\, (N-N_{_{CS}}) \Big) = \frac{2\pi}{L} \big(N - N_{_{CS}}\big)^2 
\label{ENA}
\eea
so that 
\vspace{-9mm}
\be
:\!H_f (A)\!: \big\vert N;A\big\rag = E_{N;A} \big\vert N;A\big\rag =  \frac{\pi}{2L} \, \tQ^2 \, \big\vert N;A\big\rag 
\label{Hf0}
\vspace{-2mm}
\ee
and therefore (\ref{NCS}) and (\ref{Q5shift}) imply 
\vspace{-2mm}
\be
\frac{d\tQ}{dt} = -2\,\frac{d N_{_{CS}}}{dt} = - \frac{\dot A}{\pi} L = \int_0^L dx\, \cA_2
\vspace{-2mm}
\ee
as required by the axial anomaly (\ref{anom2}).

Now the full Hamiltonian (\ref{SchwHam}) can be decomposed into its spatially constant zero mode part $H_0$, and its non-zero 
mode part which is composed of the bosonized field operators $\bar\phi_+ -  \bar\phi_-$ of (\ref{andef})-(\ref{phiposn}), 
normal ordered in $a^\dag_n, a_n$, equivalent to the $^*\!F$ independent part of the bosonic $\c$ field effective action of (\ref{fl2d}). 
This bosonized effective action describes fermion/anti-fermion pair excitations of the $\vert N;A\rag$ base states with higher energy, 
but which are otherwise independent of $A$. The bosonized Hamiltonian together with the Coulomb interaction (\ref{Hc}) can be 
diagonalized by a unitary canonical transformation $e^{-iS}$ whose net effect is to endow the boson field with a mass $M= e/\sqrt{\pi}$, 
and normal order its Hamiltonian with respect to this mass~\cite{Hetrick:1988yg,Link:1990bc,Grosse:1996hp,Hosotani:1998za}. 
This entire $n\neq 0$ bosonic part of $H$ commutes with the zero mode part $H_0$. 

Thus to find the ground state of the full Hamiltonian, one can focus only upon the zero mode subspace with no bosonic excitations, or
Coulomb term, spanned by the $\vert N;A\rag$ states, constructed in (\ref{mu5Astate}). The Hamiltonian in this spatially constant zero mode 
sector is 
\bea
H_0(\h_0;A) &=& \frac{e^2}{2L} \left(P_A + \sdfrac{ \th L}{2 \pi}\right)^2 + \frac{\pi}{2L} \bigg(P_{\h_0}- \sdfrac{AL}{\pi}\bigg)^2
= \frac{e^2}{2L} \left(\!-i \sdfrac{\pa}{\pa A} + \sdfrac{ \th L}{2 \pi}\right)^2 + \frac{\pi}{2L} \, \tQ^2 \nn
&=&-\frac{M^2L}{8 \pi} \left(\sdfrac{\!\!\!\!\pa}{\pa N_{_{CS}}}+ i\,\th \right)^2 + \frac{2\pi}{L} \, \left(\!-\sdfrac{i}{2} \sdfrac{\pa}{\pa \h_0} -N_{_{CS}}\right)^2
\label{H0}
\eea
from (\ref{Ndeta}) and (\ref{Q5shift})-(\ref{Hf0}). This Hamiltonian of two coupled degrees of freedom, $\h_0$ from the fermion sector and 
$A$ (equivalently $N_{_{CS}}$) from the gauge boson sector, may also be obtained directly from the effective Lagrangian of (\ref{fl2d}) by substituting 
$\c = -\h/\pi$ from (\ref{chietaphi}), integrating the interaction term $\int dt\, \h\, ^{*\!\!}F = -\!\int dt\, \h \dot A \rightarrow \int dt\, \dot \h A$ 
by parts, and then following standard methods.

The Hamiltonian (\ref{H0}) is left invariant under the topologically non-trivial large gauge transformation of shifting $N_{_{CS}}$ by one unit,
while at the same time shifting the phase of the fermion wave function in (\ref{Ndeta}) by $e^{2i \h_0}$, and this is explicit in the
$\vert N;A\rag$ basis, whose energy $E_{N;A}$ (\ref{ENA}) is invariant under the simultaneous shift of $N$ and $N_{_{CS}}$ by one unit.
However, because of the kinetic term in $P_A$, these fixed $\vert N;A\rag$ states do not diagonalize the full zero mode
Hamiltonian $H_0$ for any $e\!\neq\!0$. Diagonalizing $H_0$ requires instead a superposition of these states. Since 
in each sector labeled by $P_{\h_0} = 2N$, $H_0$ in (\ref{H0}) is the Hamiltonian of a simple harmonic oscillator in the 
remaining $N_{_{CS}}$ variable, the ground state wave functional is the Gaussian weighted sum~\cite{Hetrick:1988yg,Link:1990bc}  
\vspace{-8mm}
\be
\big\vert \th\big\rag = \left(\sdfrac{4}{\pi^2 ML}\right)^{\!\frac{1}{4}} \sum_{N=-\infty}^{\infty} 
\exp\left\{ - \sdfrac{2 \pi}{M  L}\,\big(N-N_{_{CS}}\big)^2  + i \th\, \big(N-N_{_{CS}}\big) \right\} e^{-iS}\, \big\vert N; A\big\rag
\label{thetavac}
\vspace{-3mm}
\ee
where the operator $e^{-iS}$ needed to diagonalize the Coulomb interaction term $H_c$ is given explicitly in 
Refs.~\cite{Link:1990bc,Grosse:1996hp,Hosotani:1998za}. The wave function of (\ref{thetavac}) is
\vspace{-2mm}
\be
\big\lag \h_0, A\big\vert \th\big\rag = \left(\sdfrac{4}{\pi^2 ML}\right)^{\!\frac{1}{4}}\sum_{N=-\infty}^{\infty} 
\exp\left\{ - \sdfrac{2 \pi}{M  L}\,\big(N-N_{_{CS}}\big)^2  + 2 i \h_0 N + i \th \,\big(N- N_{_{CS}}\big)\right\}
\label{thetawfn}
\vspace{-2mm}
\ee
in the $(\h_0, N_{_{CS}})$ coordinate basis, and $\vert \th\rag$ is an eigenstate of $H_0$, (\ref{H0}) 
with eigenvalue $M/2$.

The phase of $\vert \th \rag$ is chosen so that the exponent in (\ref{thetavac}) depends only upon the difference 
$N\!-\!N_{_{CS}}$, insuring that a shift of $N$ in the fermion sector is linked to the shift of $N_{_{CS}}$ 
in the gauge sector. Recalling (\ref{U0def})-(\ref{Uraise}), the $\vert \th \rag$ state is also an eigenstate of the unitary operator
\vspace{-2mm} 
\be
U \equiv U_0\, \exp\left( -\sdfrac{2 \pi i}{L} \,P_A \right) =  \exp\left(2 i \hat \h_0 - \sdfrac{\!\!d}{dN_{_{CS}}}\right) \qquad {\rm and} \qquad 
U \big\vert\th\big\rag =  \big\vert\th\big\rag
\ee
with unit eigenvalue. This follows from the fact that the $U_0$ part of $U$ raises $N \rightarrow N + 1$ on the $\h_0$ dependence of (\ref{thetawfn}),
while the $d/dN_{_{CS}}$ part of $U$ lowers $N_{_{CS}}\!\rightarrow\!N_{_{CS}}\!-\!1$, then relabeling the summation index in (\ref{thetawfn}) 
$N\!\rightarrow\! N\!-\!1$. Since $U$ generates topologically non-trivial gauge transformations, this proves that the $\vert\th\rag$ state is 
{\it fully gauge invariant}, as it should be -- without any phase factor which appears in some earlier treatments~\cite{Hetrick:1988yg}. 

The necessity of introducing the superposition (\ref{thetavac}) when the coupling to the gauge field is considered,
no matter how small, makes the $e\!\rightarrow \!0$ limit subtle, and is responsible for breaking the chiral symmetry. 
Indeed the action of a chiral rotation is
\vspace{-2mm}
\be
\exp\big( i \a \tQ\big)\big\vert \th \big \rag = \big\vert \th + 2 \a \big \rag 
\label{chiralrot}
\vspace{-2mm}
\ee
so $\th$ is a chiral phase and chiral symmetry is broken for any definite $\th$. 
Because the operation of $U$ on $\vert \th\rag$ generates a shift in $N$ and $N_{_{CS}}$ which is a gauge copy of the same state, 
the normalization of the wave function is determined with respect to the integration measure $\int_0^1 dN_{_{CS}}\int_0^\pi\! d \h_0$,
$N_{_{CS}}$ being integrated over only a single unit fundamental interval, {\it i.e.}
\bea
\big\lag\th\big\vert\th\big\rag  &=&\int_0^1\! d N_{_{CS}} \int_0^\pi\! d \h_0\  
\Big\vert\big\lag \h_0; A\big\vert \th\big\rag\Big\vert^2 =  
\left(\sdfrac{4}{ML}\right)^{\!\frac{1}{2}} \int_0^1\! d N_{_{CS}} \sum_{N=-\infty}^{\infty}
\exp\left\{ - \sdfrac{4 \pi}{M  L}\,\big(N-N_{_{CS}}\big)^2  \right\}\nn
&=& 2 \left(ML\right)^{\!-\frac{1}{2}}\int_{-\infty}^{\infty}\! d N_{_{CS}}\exp\left( - \sdfrac{4 \pi}{M  L}\,N_{_{CS}}^2  \right) = 1
\label{normaliz}
\vspace{-2mm}
\eea
so that after a change variables within each term of the $N$ sum in (\ref{normaliz}), the sum over $N$
can be traded for a single term, with the integration over the full real line of $N_{_{CS}}$ in the last step of (\ref{normaliz}). 

Since the fermion bilinears (\ref{bilin}) raise or lower the $\vert N\rag$ state by one unit, (\ref{condN}) their matrix elements in the
$\vert \th \rag$ state acquire a phase $e^{\pm i \th}$, and in particular
\vspace{-2mm}
\be
\big\lag\th\big\vert  \psi^\dag\,_{\!\!\!+}\psi_- \big\vert \th \big\rag= e^{-i\th} \, {\cal B}(M, L)\, \exp\left(\!-\sdfrac{\pi}{ML}\right)
\label{condML}
\vspace{-2mm}
\ee
where ${\cal B}(M, L)$ results from the normal ordering transformation $e^{-iS}$ from zero to finite boson mass $M$. This function is given by
\cite{Hetrick:1988yg,Link:1990bc,Grosse:1996hp,Hosotani:1998za}
\be
{\cal B}(M,L) = \sdfrac{M}{4 \pi}\, \exp\left\{\g_{_E} + \sdfrac{\pi}{ML} + 2 \int_0^\infty \,dx\, \left(1- e^{ML \cosh x}\right)^{-1} \right\}
\ee
where $\g_{_E} = 0.5772\dots$ is the Euler-Mascheroni constant, and it has the limiting values
\be
{\cal B}(M,L) \rightarrow \left\{ \begin{array} {lc} \frac{1}{L}\,, &ML \rightarrow 0\\ \frac{M }{4 \pi}\,\exp (\g_{_E})\,, &ML \rightarrow \infty \end{array}\right.
\ee

It is now evident that the dependence of the Gaussian width in (\ref{thetavac}) upon the product $ML$ means that the limits $e\rightarrow 0$ 
and $L \rightarrow \infty$ {\it do not commute}. In the first limit $ML \rightarrow 0$, the Gaussian becomes infinitely sharply peaked 
around $\tQ=0$ and an eigenstate of chiral charge. Hence the chiral symmetry is restored, all $\th$ dependence drops out, and both
\vspace{-2mm}
\be
z_0 = \big\lag \th\big\vert e^{2 i \hat\h_0} \big\vert \th \big\rag = \exp\left(\!-\sdfrac{\pi}{ML}\right) 
\label{z0expect}
\vspace{-2mm}
\ee
and the chiral symmetry breaking condensate (\ref{condML}) vanish exponentially as $ML \rightarrow 0$. 
The Goldstone pole decouples completely, as in the strictly free fermion theory with $e\! \equiv\!0$ exactly as before.

On the other hand if we take $L\!\rightarrow \!\infty$ with $e$ fixed, the anomaly source term in (\ref{anom2}) cannot be neglected, as was 
assumed in (\ref{divJ5}). The would-be Nambu-Goldstone excitation $\c = -\h/\pi$ combines with the electric field through the Gauss Law 
constraint (\ref{Gauss}), solved by $E_x= e^2\c + E_0$ (where $E_0$ is an integration constant, a spacetime constant background electric field), 
and $\c$ becomes massive with $M^2= e^2/\pi$. In other words, the classically constrained gauge field is `eaten' by the propagating would-be 
Goldstone boson, just as in spontaneous symmetry breaking by the Stueckelberg-Higgs mechanism. Nevertheless it should be 
emphasized that the chiral symmetry and chiral Ward Identities are actually broken {\it explicitly} by the anomaly, unlike the more familiar
case of spontaneous symmetry breaking by a scalar field expectation value, where the Ward identities of the symmetry are
non-anomalous and preserved.

The interesting intermediate case is the limit $e\rightarrow 0, L\rightarrow \infty$ with $eL$ {\it fixed}. All the $N$ in (\ref{thetavac}) still contribute, 
and the condensates (\ref{condML}) and (\ref{z0expect}) remain non-vanishing. Chiral symmetry remains broken and $z_0 \neq 0$ even for infinitesimally small $e$.
In this limit the conditions for the proof of Goldstone's theorem of Sec.~\ref{Sec:NambGold} are satisfied, and the massless Goldstone pole exists
over the range of distance or momentum scales 
\vspace{-2mm}
\be
|x| \ll e^{-1} \lesssim L    \quad {\rm or} \quad k \gg e \gtrsim \sdfrac{1}{L}
\label{xkrange}
\vspace{-2mm}
\ee
which becomes arbitrarily large as $e\rightarrow 0$ and $L \rightarrow \infty$ together. This is the case of spontaneous symmetry
breaking, with the Goldstone mode appearing through an arbitrarily weak attractive interaction between the fermion/anti-fermion
pairs, similar to that in formation of Cooper pairs in condensed matter systems. Further verification of this result is obtained by 
computing the condensate correlation function \cite{Hetrick:1988yg,Hosotani:1998za}
\vspace{-4mm}
\bea
&&\hspace{1cm} \big\lag\th\big\vert \psi^\dag\,_{\!\!\!+} \psi_-(x)\  \psi^\dag\,_{\!\!\!+} \psi_-(0) \big\vert \th \big\rag =
 \big[{\cal B}(M,L)\big]^2 \exp \left\{\sum_{n\neq 0}^\infty \frac{e^{ik_n x}}{\left[n^2 + (ML/2\pi)^2\right]^\frac{1}{2}}\right\}\label{condcor}\\
&&\rightarrow \big[{\cal B}(M,L)\big]^2 \exp \left\{\sum_{n\neq0} \frac{e^{2 \pi inx/L}}{|n|}  \right\}\rightarrow
 \left(\frac{M }{4 \pi}\,e^{\g_{_E}}\right)^2  \exp \left\{ - 2 \ln \left(\sdfrac{2 \pi |x|}{L} \right) \right\} 
 =  \left(\frac{M }{4 \pi}\,e^{\g_{_E}}\right)^2\! \left(\frac{L}{2 \pi x}\right)^2\nonumber
\eea
in the limit $|x| \ll e^{-1} \sim L $, since $n\sim L/|x| \gg ML$ dominate the sum. The argument of the exponential in (\ref{condcor})
is the commutator function of the massless $\h$ field in (\ref{chietaphi})-(\ref{phiposn}), {\it i.e.} the superfluid Nambu-Goldstone mode. There 
is {\it `quasi-long range order'} for this range of $|x|$, which decreases as a power due to the logarithmic 
massless boson propagator in $D\!=\!2$ \cite{Kleinert:1989}. 	

The $\vert\th\rag$ state is easily generalized to a state with finite chiral density at zero temperature. The Grand Potential
$\W = H_0 - \tm \tQ$ in the zero mode sector suggests shifting $N\!- \!N_{_{CS}} \rightarrow N \!- \!N_{_{CS}}\! - \!\tm L/2\pi$ in the Gaussian
(\ref{thetawfn}). However shifting the centroid of a Gaussian away from $\tQ\!=\!0$ yields a time dependent state. The time 
dependent Gaussian wave function 
\bea
&&\hspace{-5mm} \Psi_{\th;\tm}(\h_0,A; t) = \big\lag \h_0,A \big\vert e^{-i H_0 t}\big\vert  \th; \tm\big\rag 
=\left(\sdfrac{4}{\pi^2 ML}\right)^{\!\frac{1}{4}}\!\! \sum_{N=-\infty}^{\infty}\! \!
\exp\left\{ - \sdfrac{2 \pi}{M  L}\Big(N-N_{_{CS}} + \overline N_{_{CS}}(t) \Big)^2   \right.\nn
&&\hspace{1cm}\bigg.  + 2 i\h_0 N  + i \th\, \big(N-N_{_{CS}}\big) 
- \sdfrac{2 \pi i }{M^2L} \,\dot{\overline N}_{CS}(t)\Big(2 N-2 N_{_{CS}}+ \overline N_{_{CS}}(t)\Big) - \sdfrac{iMt}{2} \bigg\} 
\label{mu5vac}
\vspace{-2mm}
\eea
satisfies the time dependent Schr\"odinger eq. $i\pa_t\Psi  = H_0\Psi$, provided $\overline N_{_{CS}}(t)$ satisfies \cite{Habib:1995ee}
\vspace{-2mm}
\be
\left (\sdfrac{d^2}{dt^2} + M^2\right) {\overline N}_{CS}= 0\,,\qquad \overline N_{_{CS}}(t) = -\sdfrac{\tm L}{2 \pi}\, \cos \,(Mt)
\label{classNCS}
\vspace{-2mm}
\ee
{\it i.e.} $\overline N_{_{CS}}(t)$ is the particular solution of the classical oscillator eqs. of motion following from $H_0$
satisfying the initial conditions $\overline N_{_{CS}}(0) = -\tm L/2\pi$ and $\dot{\overline N}_{CS}(0)=0$.

The chiral symmetry is broken and the conditions for the Goldstone theorem are again satisfied in the same sense as before 
for the coordinate or momentum ranges (\ref{xkrange}). The expectation value of the chiral symmetry breaking condensate 
operator $\psi^\dag\,_{\!\!\!+}\psi_- $ is 
\vspace{-2mm}
\be
\big\lag \th;\tm\big\vert e^{i H_0 t} \,\psi^\dag\,_{\!\!\!+}\psi_-  \,e^{-i H_0 t}\big\vert \th; \tm\big\rag =
\exp\Big\{\sdfrac{2i \tm}{M} \sin(Mt)\Big\} \,e^{-i\th} \,e^{ - \pi/ML} \, {\cal B}(M, L)
\label{condmu}
\vspace{-2mm}
\ee
in place of (\ref{condML}), in the time dependent state of non-zero chemical potential $\tm$. Recalling (\ref{chiralrot}), and 
with $\sin(Mt) \approx Mt$ for $Mt \ll 1$, (\ref{condmu}) shows that the chiral phase angle rotates with constant positive speed for times $t \ll M^{-1}$, 
as expected for the phase of a condensate with $\dot \h = \tm$ and $\tQ = 2$.
The time range $t \ll M^{-1}$ becomes infinitely long as $M\!\rightarrow\! 0$, consistent with the spatial range of
(\ref{xkrange}) where the Goldstone theorem of Sec.~\ref{Sec:NambGold} applies.

\end{document}